\documentclass[a4paper,10pt,twoside]{cpc-hepnp}
\usepackage{CJK,upgreek,fancyhdr}
\usepackage{multicol,color}
\usepackage{graphicx,subfigure,dcolumn}
\usepackage{epsfig}
\usepackage{booktabs,slashed}
\usepackage{amssymb,bm,mathrsfs,bbm,amscd,amsfonts}
\usepackage[tbtags]{amsmath}
\usepackage{lastpage}
\usepackage{hyperref}

\newcommand{\f}{\begin{equation}}
\newcommand{\ff}{\end{equation}}
\newcommand{\fa}{\begin{eqnarray}}
\newcommand{\ffa}{\end{eqnarray}}

\providecommand{\U}[1]{\protect\rule{.1in}{.1in}}
\begin{document}
\title{Power Law of Shear Viscosity in Einstein-Maxwell-Dilaton-Axion model
\thanks{Supported by
National Natural Science Foundation of China under Grant (Nos.11275208 and 11575195),
Opening Project of Shanghai Key Laboratory of High Temperature
Superconductors (No. 14DZ2260700),
Jiangxi young scientists (JingGang Star) program and 555 talent project of Jiangxi Province.}}

\author{Yi Ling $^{1,2,3}$ \email{lingy@ihep.ac.cn}
\quad Zhuoyu Xian $^{1}$ \email{xianzy@ihep.ac.cn}
\quad Zhenhua Zhou $^{1}$ \email{zhouzh@ihep.ac.cn}}

\maketitle

\address{
$^1$ Institute of High Energy Physics, Chinese Academy of Sciences, Beijing 100049, China\ \\
$^2$ Shanghai Key Laboratory of High Temperature Superconductors, Shanghai, 200444, China\ \\
$^3$ School of Physics, University of Chinese Academy of Sciences, Beijing 100049, China}
\maketitle

\begin{abstract}
We construct charged black hole solutions with hyperscaling violation in the infrared(IR) region in Einstein-Maxwell-Dilaton-Axion theory and investigate the temperature behavior of the ratio of holographic shear viscosity to the entropy density.  When translational symmetry breaking is relevant in the IR, the power law of the ratio is testified numerically at low temperature $T$, namely, $\eta/s\sim T^\kappa$, where the values of exponent $\kappa$ coincide with the analytical results. We also find that the exponent $\kappa$ is not affected by irrelevant current, but is reduced by the relevant current.
\end{abstract}



\section{Introduction}

In last decade the development of holographic techniques has given
new insights into the hydrodynamics with strong couplings.
One celebrated achievement is the Kovtun-Son-Starinets (KSS) bound
for the ratio of shear viscosity to entropy density
\cite{Kovtun:2004de}
\begin{equation}\label{constantbound}
\frac{\eta}{s}\geq \frac1{4\pi},
\end{equation}
which is considered as a fundamental bound for near perfect fluid
with strong interactions. However, in recent years
counter-examples which violate the KSS bound (\ref{constantbound})
have been found in holographic literature, including the higher
derivative gravity \cite{Brigante:2007nu,Kats:2007mq}, anisotropic
system \cite{Rebhan:2011vd,Ge:2014aza,Mateos:2011ix,Mateos:2011tv}
as well as isotropic system without translational invariance
\cite{Hartnoll:2016tri,Burikham:2016roo,Liu:2016njg,Alberte:2016xja,Davison:2014lua,Ling:2016ien,Wang:2016vmm,Garcia-Garcia:2016hsd}.
In the latter case, shear viscosity loses its
hydrodynamical interpretation because of the non-conservation of
momentum and is usually defined by Kubo Formula
\begin{equation}\label{viscocitydef}
\eta=\lim_{\omega\to 0}\frac1{\omega}\text{Im}
G^R_{\hat{T}^{xy}\hat{T}^{xy}} (\omega,k=0),
\end{equation}
where $G^R_{\hat{T}^{xy}\hat{T}^{xy}}$ is the retarded Green
function of the energy-momentum tensor operator
$\hat{T}^{xy}$ in the dual boundary theory. Nevertheless,
it is still quite instructive to investigate the temperature
behavior of the ratio of shear viscosity to entropy density in
general holographic models without translational
invariance.

Historically, the breaking of translational invariance is
introduced in holography to study the transportation of the
dual system with momentum dissipation
\cite{Horowitz:2012gs,Horowitz:2012ky,Horowitz:2013jaa,Andrade:2013gsa,Vegh:2013sk,Hartnoll:2012rj,Ling:2013nxa,Blake:2013owa,Kiritsis:2015oxa,Davison:2014lua,Ling:2014laa,Blake:2014yla,Baggioli:2014roa,Zeng:2014uoa,Amoretti:2014zha,Davison:2013txa}. In this setup, new geometries in the IR may emerge and often
accompany with new scaling relations, leading to new scaling
behavior of thermodynamic quantities or Green functions with
temperature $T$ or frequency
$\omega$ \cite{Hartnoll:2015rza,Hartnoll:2015faa,Hartnoll:2014cua,Donos:2013eha,Donos:2014oha,Donos:2014uba,Donos:2012ra,Gouteraux:2014hca,Donos:2012js}.

In particular, by virtue of recent progress in
\cite{Hartnoll:2016tri} people have learned that when the
translation symmetry breaking is relevant in the far IR,
the ratio exhibits a power law behavior with the temperature
\begin{equation}\label{etapowerlaw}
\frac{\eta}{s}\sim T^\kappa,
\end{equation}
which reflects the scaling symmetry emerging in the IR. Moreover,
a new bound for the exponent $\kappa$ was proposed as $\kappa\leq
2$ there, which might be supported by a heuristic argument
based on the bound for the rate of entropy production,
\begin{equation}\label{sbound}
\frac{\hbar}{k_BT}\frac{d\log(s)}{dt}\gtrsim 1.
\end{equation}
Other holographic models are investigated in
\cite{Wang:2016vmm,Liu:2016njg,Burikham:2016roo,Alberte:2016xja,Davison:2014lua,Garcia-Garcia:2016hsd},
which also satisfy the bound $\kappa\leq 2$. But soon later in
\cite{Ling:2016ien}, we find $\kappa> 2$ is possible when the
scaling of Lifshitz
\cite{Kachru:2008yh,Taylor:2008tg,Gubser:2009cg} or hyperscaling
violation
\cite{Gubser:2009qt,Cadoni:2009xm,Goldstein:2009cv,Charmousis:2010zz,Gouteraux:2012yr,Perlmutter:2010qu,Gouteraux:2011ce,Iizuka:2011hg,Ogawa:2011bz,Huijse:2011ef,Alishahiha:2012cm,Bhattacharya:2012zu,Kiritsis:2015oxa,Gouteraux:2011qh,Dong:2012se,Gouteraux:2013oca}
emerges in the IR. While the bound for the rate of entropy
production (\ref{sbound}) is not violated, since the origin of
the power law (\ref{etapowerlaw}) should be understood as
the nontrivial anomalous dimensions of (\ref{viscocitydef}) under
the rescaling of the IR solution
\cite{Hartnoll:2016tri,Ling:2016ien}.

Furthermore, in \cite{Ling:2016ien} we have analytically derived a
formula for $\kappa$ in a general Einstein-Maxwell-Dilaton-Axion (EMD-Axion) model with spatial
dimension $d$, dynamical critical exponent $z$ as well as
hyperscaling violating exponent $\theta$. Specifically, we find
\begin{equation}\label{scaleformula}
\frac{\eta}{s}\sim T^{\frac{d+z-\theta}{z} \left(-1+\sqrt{\frac{8
(z-1)}{(d+z-\theta) (1+e^2)}+1}\right)},
\end{equation} where
$e^2$ is defined as the ratio of Maxwell term and one of the
lattice terms in the Lagrangian. However, in \cite{Ling:2016ien}
only for the case of $e^2=0$ has this formula (\ref{scaleformula})
been justified by numerical calculation on neutral background. In
this paper we intend to continuously testify the validity of
(\ref{scaleformula}) on charged background within
EMD-Axion models. Schematically, the case of
$e^2\neq0$ can be realized by relevant currents. We
will numerically construct specific charged backgrounds with
ultraviolet(UV) completion and then compute the power law of
the ratio of shear viscosity to entropy density. As a result, we
will show that the formula for the ratio $\eta/s$ in
(\ref{scaleformula}) which was previously proposed in
\cite{Ling:2016ien} based on the scaling analysis can be
justified even for $e^2\neq 0$ indeed.

In this paper, we adopt the statement about `(marginally) relevant' and `irrelevant' in \cite{Gouteraux:2014hca}: the current or the axion being
(marginally) relevant means that the Maxwell or the axion terms is the same order as the curvature term and the dilaton
potential in Lagrange in the power of the radial coordinate; the current or the axion being irrelevant means that the Maxwell or the axion terms is subleading comparing to the curvature term and the dilaton potential. Roughly speaking, a field being (marginally) relevant or irrelevant depends on whether or not it would deform the IR geometry strongly.

\section{EMD-Axion model and hyperscaling violating metric}
In this section we will present our holographic setup and then
outline the logic leading to the formula for the exponent $\kappa$
in (\ref{scaleformula}) based on the scaling analysis. The action
of a general EMD-Axion model in $d+2$ space time reads
 as
\begin{equation}\label{action}
\mathcal{S}=\int dt d^dx dr\sqrt{-g}( R+ \mathcal{L}_m) ,\quad
\mathcal{L}_m = -\frac12(\partial \phi
)^2-\frac{J(\phi)}{2}\sum_{i=1}^d(\partial\chi_i)^2+V(\phi)-\frac{Z(\phi)}{4}F^2,
\end{equation}
where $\chi_i(i=1,2,\cdots,d)$ are axions and $J(\phi),Z(\phi),
V(\phi)$ are coupling functions or potential of the dilaton field
$\phi$. Given above action, the equations of motion can be derived
as follows
\begin{subequations}\label{EOM}\begin{align}
&
R_{\mu \nu }+ \frac1{d}{g_{\mu \nu }T}-T_{\mu \nu }=0,\quad T_{\mu
\nu
}=-\frac1{\sqrt{-g}}\frac{\delta(\sqrt{-g}\mathcal{L}_m)}{\delta
g^{\mu \nu }} =\frac12 g_{\mu \nu } \mathcal{L}_m -
\frac{\delta\mathcal{L}_m}{\delta g^{\mu \nu }},    \label{EE}
\\&
\nabla^2\phi-\frac{J'(\phi)}{2}\sum_{i=1}^d(\partial\chi_i)^2+V'(\phi)-\frac{Z'(\phi)}{4}F^2=0,
\\&
\nabla^\nu(Z(\phi)F_{\mu\nu})=0,    \label{ME}
\\&
\nabla^\mu(J(\phi)\partial_\mu\chi_i)=0, \quad i=1,2,\cdots,d.
\end{align}\end{subequations}
For simplicity, we only consider isotropic solutions with
following ansatz
\begin{equation}\begin{split}\label{metricandemtensor}
&
ds^2=-g_{tt}(r)dt^2+g_{rr}(r)dr^2+g_{xx}(r)\sum_{i=1}^d dx_i^2,
\\&
\phi=\phi(r),\quad \chi_i=kx_i,\quad A=A_t(r)dt,
\end{split}\end{equation}
where translational invariance is broken by the axions $\chi_i$
but the metric and energy-momentum tensor remain to be
homogeneous.

As stated above, we are interested in solutions interpolating
between AdS in the UV and hyperscaling violating solution in the
IR. It can be realized by the process of UV completion
\cite{Charmousis:2010zz}. Firstly, we construct a hyperscaling
violation solution with running dilaton. Secondly, we modify
the local behaviors of potentials to graft the solution onto AdS
in the UV. Generally, solutions with AdS exist when the
dilaton reaches the extremal point of its potential and the
Lorentz symmetry is maintained \cite{Gouteraux:2012yr}.

Let us find hyperscaling violating solutions firstly. When
the potentials behave as
\begin{equation}\label{potentials}
V(\phi)\sim V_0 e^{\alpha\phi}, \quad J(\phi)\sim e^{\beta\phi}, \quad Z(\phi)\sim e^{\gamma\phi},\quad \text{when } \phi\to\pm\infty,
\end{equation}
it is found in \cite{Gouteraux:2014hca,Donos:2014uba} that
scaling solutions with hyperscaling violation exist, whose metric
and matter fields read as
\begin{equation}\label{HV}
ds^2= r^\frac{2\theta}{d} \left( -\frac{dt^2}{r^{2z}}+
\frac{L^2 dr^2}{r^2}+\frac{\sum_{i=1}^d dx_i^2}{r^2}
\right),\quad A=Qr^{\zeta-z}dt,\quad e^\phi=r^{\epsilon},\quad
\chi_i=kx_i,
\end{equation}
where $k$ characterizes the scale of breaking of translational invariance and $\zeta$ is called conduction exponent \cite{Gouteraux:2013oca}. Parameters $\{z,\theta,\zeta,\epsilon,k,Q\}$ in ansatz (\ref{HV}) are determined by equations of motion (\ref{EOM}). The solutions are found to be classified into four classes \cite{Gouteraux:2014hca}, depending on the parameters $\{\alpha,\beta,\gamma,V_0\}$ in potentials (\ref{potentials}). Among them, the explicit solutions with (marginally) relevant axion are shown in Appendix \ref{SectionIRSolution}. The metric in (\ref{HV}) can be deformed into a black hole solution
\begin{equation}\label{HVBH}
ds^2= r^\frac{2\theta}{d} \left( -\frac{f(r)dt^2}{r^{2z}}+
\frac{L^2 dr^2}{r^2 f(r)}+\frac{\sum_{i=1}^d dx_i^2}{r^2}
\right),
\end{equation}
where the blackness function is
\begin{equation}
f(r)=1-\left(\frac{r}{r_+}\right)^{\delta_0}, \quad \delta_0=d+z-\theta.
\end{equation}
Then, the Hawking temperature and entropy density are separately
given by
\begin{equation}\label{TsHV}
T=\frac{|\delta_0|}{4\pi}r_+^{-z}, \quad s=4\pi r_+^{\theta-d}=4\pi\left(\frac{4\pi T}{|\delta_0|}\right)^\frac{d-\theta}{z}.
\end{equation}
From Maxwell equation (\ref{ME}), we have conserved charge density
\begin{equation}\label{rho}
\rho = \sqrt{-g} Z(\phi) F^{rt}.
\end{equation}
It can be shown that there exists a scaling relation
\begin{equation}
x\to c x,\,r\to c r,\,t\to c^z t,\,ds\to c^{\theta/d}ds,\,T\to c^{-z}T,\,s\to c^{d-\theta}s.
\end{equation}
We will come back to the UV completion and modify the potentials in next section.

Now we turn to the study of shear viscosity $\eta$. From
Kubo formula (\ref{viscocitydef}), $\eta$ can be derived by
perturbing $(\delta g)^x{}_y=g^{xx}\delta g_{xy}=h(r)e^{-i\omega
t}$, which is the dual field of operator $\hat{T}^{xy}$. Einstein
equations (\ref{EE}) give rise to the shear perturbation equation
\begin{equation}\label{perturh}
\frac1{\sqrt{-g}}\partial_r(\sqrt{-g}g^{rr}\partial_r h(r)) + (g^{tt}\omega^2- m(r)^2) h(r) =0,
\end{equation}
with varying mass
\begin{equation}\label{massdef}
m(r)^2= 2( g^{xx}T_{xx}-\frac{\delta T_{xy}}{\delta g_{xy}}).
\end{equation}
$h(r)$ is required to be regular at the horizon and equal to $1$
at the conformal boundary $r_\partial$. The varying mass is
supposed to satisfy the condition $m(r)^2\geq0$ in the models
considered so far. Here, we have $m(r)^2=J(\phi)k^2g^{xx}$.

$\eta/s$ can be obtained by the weaker horizon formula \cite{Lucas:2015vna,Hartnoll:2016tri}
\begin{equation}\label{viscosityh+d+2}
\frac{\eta}{s}=\frac1{4\pi}h_0(r_+)^2,
\end{equation}
where $h_0(r)$ is the solution at $\omega=0$ and $r_+$ is the location of horizon.

Following the analysis presented in \cite{Ling:2016ien}, one can
calculate the exponent $\kappa$ of $\frac{\eta}{s}\sim T^\kappa$.
Here we present a simpler derivation with the use of formula
(\ref{viscosityh+d+2}). If the axion is (marginally) relevant, by using
Einstein equations (\ref{EE}) and black hole metric (\ref{HVBH}),
we have
\begin{equation}\label{massEMDA}
m(r)^2=M^2r^{-\frac{2\theta}{d}},\quad M^2=\frac{2\delta_0(z-1)}{(1+e^2)L^2},
\end{equation}
where
\begin{equation}
e(r)^2=-\frac{Z(\phi)}{4}F^2\left/\left(\frac12J(\phi)(\partial\chi_x)^2\right)\right.\geq0,\quad \chi_x=kx.
\end{equation}
$e(r)^2$ is just the ratio of the Maxwell term to one
of the axion terms in Lagrangian (\ref{action}). It goes to a
nonzero constant in the far IR if the current is also
(marginally) relevant, otherwise it goes to zero. Thus at leading order, we just set $e(r)^2=e^2$. If the
axion is irrelevant, $m(r)^2$ goes to zero in the far IR, then we
set $M^2=0$, which is valid at leading order.

By using (\ref{massEMDA}) and metric (\ref{HVBH}), we rewrite perturbation equation (\ref{perturh}) at $\omega=0$
\begin{equation}\label{peq}
\partial_r(r^{1-\delta_0}f(r)\partial_r h_0(r))-M^2L^2 r^{-\delta_0-1}  h_0(r) =0.
\end{equation}
Solving this equation we can separately obtain the
asymptotical expansion of $h_0(r)$ near the boundary and its value
on the horizon as
\begin{equation}\label{hb}
h_0(r\to r_i)=c\left[\left(\frac{r}{r_+}\right)^{\delta_0-\delta_{\hat{T}}}+\cdots + G \left(\frac{r}{r_+}\right)^{\delta_{\hat{T}}}+\cdots\right],\quad h_0(r_+)=c H,
\end{equation}
where
\begin{equation}\label{deltaT}
\delta_{\hat{T}}=\frac{\delta_0}{2}\left(1+\sqrt{1+\left(\frac{2ML}{\delta_0}\right)^2} \right).
\end{equation}
We have abbreviated the series
$\left(\frac{r}{r_+}\right)^{\delta_0-\delta_{\hat{T}}+n\delta_0}$
($n\geq1$) following
$\left(\frac{r}{r_+}\right)^{\delta_0-\delta_{\hat{T}}}$ and the
series $\left(\frac{r}{r_+}\right)^{\delta_{\hat{T}}+m\delta_0}$
($m\geq1$) following
$\left(\frac{r}{r_+}\right)^{\delta_{\hat{T}}}$ to ellipsis.
$G$ and $H$ are some $r_+$ independent constants, which only
depend on $\delta_0$ and $\delta_{\hat{T}}$. Coefficient $c$
should be determined by the boundary condition
$h_0(r_\partial)=1$. $r_i$ is the boundary of the region where
the black hole with hyperscaling violation can be described by
(\ref{HVBH}).
When consider the second-order variation of the action
in (\ref{action}) over a fixed background with the perturbed
metric $h_0(r)$, the boundary part of the variation with the branch
of $r^{\delta_0-\delta_{\hat{T}}}$ ( $r^{\delta_{\hat{T}}}$) in
(\ref{hb}) is divergent (finite), thus the branch of
$r^{\delta_0-\delta_{\hat{T}}}$ ( $r^{\delta_{\hat{T}}}$) is
non-normalizable (normalizable). Then
$h_0(r_i)\approx
c\left(\frac{r_i}{r_+}\right)^{\delta_0-\delta_{\hat{T}}}$ is a
good approximation when the black hole is
near-extremal.

Since the UV completion is taken into account, one should be
cautious that $r_i$ is just an intermediate scale but not the
conformal boundary $r_\partial$. The region between $r_i$ and
$r_\partial$ is AdS deformed by matter fields. The boundary
condition requires $h_0(r_\partial)=1$. As explained
in \cite{Hartnoll:2016tri}, when going from $r_\partial$ to $r_i$,
$h_0(r)$ decreases monotonously from 1 to a value
$\Gamma$ when $m(r)^2>0$. We have introduced a
`tunneling rate' $\Gamma$ to characterize how $h_0(r)$
tunnels from $r_\partial$ to $r_i$. The tunneling
rate $\Gamma$ and intermediate scale $r_i$ should be independent
of $T$ when the scale of $T$ is much less than other scales, such
as the source of dilaton and $k$ in axion. It is because that
scale $T$ is not the dominating scale driving the
renormalization group(RG) flow from AdS to hyperscaling
violating solution. Scale $T$ becomes important only when we go
into the far IR. Then we have
\begin{equation}
\Gamma=h_0(r_i)\approx c\left(\frac{r_i}{r_+}\right)^{\delta_0-\delta_{\hat{T}}}.
\end{equation}
By working out $c$, we can determine the horizon value
\begin{equation}
h_0(r_+)=\Gamma H \left(\frac{r_+}{r_i}\right)^{\delta_0-\delta_{\hat{T}}}.
\end{equation}
By virtue of formula (\ref{viscosityh+d+2}), we finally
obtain
\begin{equation}\label{bound}
\frac{\eta}{s}=\frac{\Gamma^2 H^2}{4\pi}\left(\frac{r_+}{r_i}\right)^{2(\delta_0-\delta_{\hat{T}})}\propto T^{\frac{d-\theta+z}{z}\left(-1+\sqrt{1+\frac{8 (z-1)}{(d-\theta+z)(1+e^2)}}\right)}, \text{ when } T\to0,
\end{equation}
where (\ref{TsHV}) and (\ref{deltaT}) have been used. One
can also employ UV-IR matching
\cite{Donos:2012ra,Faulkner:2009wj,Faulkner:2011tm} to reproduce
the same result as what is performed in \cite{Ling:2016ien}. As
one can see, when axion is absent or irrelevant, we have
$m(r)^2=0$ at leading order, then $\eta/s\sim T^0$
\cite{Kuang:2015mlf,Kolekar:2016pnr}. Next we will numerically
testify this formula when $e^2=0$ or $e^2\neq0$.

\section{UV completion and numerical results}
In this section we specifically construct the background
interpolating hyperscaling violating solution (\ref{HV}) in the IR
and AdS solution in the UV with finite temperature and charge
density. On one hand, as explained in Appendix
\ref{SectionIRSolution} or in \cite{Gouteraux:2014hca},
solution (\ref{HV}) can be constructed by choosing exponential
potentials (\ref{potentials}) with running dilaton. On the other
hand, AdS can be constructed by finding extremal points of
constant dilaton with lorentz symmetry. Here we adopt dilaton
$\phi$ to interpolate the UV and the IR solutions. It requires
some special settings for the potential $V(\phi)$.

\subsection{UV completion}
In this paper, we focus only on $\theta<d$ since in this region
the entanglement entropy obeys area to volume law, which is
considered as normal behavior of QFT
\cite{Ling:2016ien,Dong:2012se}. In this situation the location of
IR is $r\to+\infty $. Following the discussion in
\cite{Ling:2016ien,Gouteraux:2012yr}, the constraints about
$(z,\theta)$ are reduced to
\begin{equation}\label{constraint}
(\theta \leq 0\land z > 1)\lor \left(0<\theta < d\land z > \frac{\theta }{d}+1\right).
\end{equation}
Then it leads to $\delta_0>0$. We have excluded the
two cases of $\theta=d$ and $z=\frac{\theta }{d}+1$ which
can not be reached by running dilaton, as shown in Appendix
(\ref{SectionIRSolution}) or \cite{Gouteraux:2014hca}. We choose
the branch of $\phi\geq0$. From the requirement of the potentials
(\ref{potentials}), one can see that our solution can flow to
hyperscaling violating solution in the IR if $\phi\to+\infty
(r\to+\infty)$. It requires $\epsilon>0$ in (\ref{HV}). We conduct
the UV completion by modifying the potential $V(\phi)$ but fixing
the other two coupling potentials
\begin{equation}
J(\phi)= e^{\beta\phi}, \quad Z(\phi)= e^{\gamma\phi}.
\end{equation}
From the analysis in Appendix (\ref{SectionIRSolution}), we find a
universal behavior of $V(\phi)\sim r^{-\frac{2\theta}{d}}$ in the
coordinate of ansatz (\ref{HV}). So, when approaching the UV
($r\to 0$), the qualitative behavior of $V(\phi)$
depends on the sign of $\theta$. On the side of the UV, AdS is
allowable if axion and gauge field are turned off and
$\phi=\phi_*$ is an extremal point of $V(\phi)$, where $V(\phi_*)$
stands for cosmological constant. Without loss of generality, we
choose $\phi_*=0$. Then a realistic strategy is modifying
$V(\phi)$ as
\begin{equation}\label{V}
V(\phi)=\left\{\begin{array}{lll}
\frac{2 d}{\alpha ^2}\sinh ^2\left(\frac{\alpha  \phi }{2}\right)+(d+1) d,& V_0=\frac{d}{2\alpha^2},& \text{ for } \theta<0 \\
(d+1) d,   & V_0=(d+1) d , & \text{ for } \theta=0 \\
\left(d (d+1)-\frac{V_0}{2}\right) \left(1-\tanh ^2(\alpha  \phi )\right)+\frac{V_0}{2 \cosh (\alpha  \phi )}, & V_0=2 d \left(\frac{1}{\alpha ^2}+2 d+2\right) ,& \text{ for } d>\theta>0
\end{array}\right..
\end{equation}
In Appendix \ref{SectionIRSolution}, we have $\alpha  \epsilon =-\frac{2 \theta }{d}$. Then $\alpha$ has the opposite sign of $\theta$.
So, as one can see, each $V(\phi)$ approaches $V_0e^{\alpha\phi}$ when
$\phi\to+\infty$ and approaches $d(d+1)$ when $\phi\to0$. Without loss of generality, we have chosen the AdS radius to be 1.
However, the intermediate behavior is not very important.

$AdS_{d+2}$ vacuum is always allowable. We choose the first type
quantization, and the scaling dimensions about the dual source of
dilaton $\phi^{(0)}$ and operator $\mathcal{O}_\phi$ are
determined by the small $\phi$ expansion of $V(\phi)$. When
$\theta=0$, as dilaton is massless, we have
$\Delta_{\phi^{(0)}}=0$ and $\Delta_{\mathcal{O}_\phi}=d+1$, which
is marginal deformation. We expect it to be marginally relevant to
drive the solution away from AdS, as $\phi=0$ is not a stable
point when axion and gauge field are turned on. When
$\theta\neq0$, as $V(\phi)=d(d+1)+\frac{d}{2}\phi^2+\cdots$,  we
have $\Delta_{\phi^{(0)}}=1$ and $\Delta_{\mathcal{O}_\phi}=d$,
which is relevant deformation.

\subsection{Numerical calculation and results}
We use the following ansatz for numerical calculation
\begin{equation}\begin{split}\label{ansatz}
&
ds^2=\frac1{u^2}\left(-(1-u)U(u)e^{-S(u)}dt^2+\frac{du^2}{(1-u)U(u)}+ \sum_{i=1}^d dx_i^2 \right),
\\&
\phi=\phi(u),\quad \chi_i=kx_i,\quad A=(1-u)A(u)dt.
\end{split}\end{equation}
The conformal boundary is located at $u=0$ while the horizon at
$u=1$. The temperature and the entropy density are
$T=\frac1{4\pi}U(1)e^{-S(1)/2}$ and $s=4\pi$.

The $AdS_{d+2}$ vacuum corresponds to $U=1,~S=\phi=A=0$.
Boundary conditions at the horizon are regular conditions.
Boundary conditions at conformal boundary should satisfy the
scaling dimensions, which depend on the potential $V(\phi)$ as
well as the value of $\theta$.

Explicitly, the asymptotic expansions near the conformal
boundary are
\begin{equation}\begin{split}
U(u)    &= 1+ \cdots + \varepsilon u^{d+1}+ \cdots, \\
e^{-S(u)}&= 1 +  \cdots,\\
A(u)    &=\mu +\cdots+\rho u^{d-1}+\cdots, \\
\phi(u) &= \begin{cases}
\lambda  +\cdots + \nu u^{d+1} + \cdots ,& \theta=0 \cr
\lambda u +\cdots + \nu u^{d} + \cdots ,& \theta\neq0 \end{cases}
\end{split}\end{equation}
where $\mu$ is the chemical potential and $e^{-S(0)}$ has been set
to $1$ by rescaling $t$. The different boundary conditions of
$\phi(u)$ come from the different choices of $V(\phi)$ in
(\ref{V}). We can work on either grand canonical ensemble or
canonical ensemble.
\subsubsection{\bf Grand canonical ensemble}

In grand canonical ensemble, we control the value of chemical
potential $\mu$.

When $\theta=0$, the boundary conditions at conformal boundary are
$U(0)=1,~S(0)=0,~A(0)=\mu,~\phi(0)=\lambda$. We work in the unit
of $k$. The dimensionless quantities parameterizing the family of
black hole solutions are
$\{\frac{T}{k},~\frac{\mu}{k},~\lambda\}$.
When $\theta\neq0$, the boundary conditions are
$U(0)=1,~S(0)=0,~A(0)=\mu,~\phi'(0)=\lambda$.  The dimensionless
quantities are
$\{\frac{T}{k},~\frac{\mu}{k},~\frac{\lambda}{k}\}$.

We numerically construct the interpolating solutions for
$\phi\geq0$. When dropping down $T/k$, we should fix the
values of $\{\frac{\mu}{k},\lambda\}$ (for $\theta=0$) or
$\{\frac{\mu}{k},\frac{\lambda}{k}\}$ (for $\theta\neq 0$)
within an appropriate region respectively, in order to reach hyperscaling violating solution in the IR at low $T/k$.

The dimensionless entropy density and charge density are $s/k^d$
and $\rho/k^d$. We calculate $\eta/s$ by using
(\ref{viscosityh+d+2}) and found $\frac{\eta}{s}\leq\frac1{4\pi}$
for all the time, because of the breaking of translational
invariance.

At high $T/k$, the scaling relation is controlled by
AdS in the UV, which gives rise to the power laws of $s\sim
T^d$ and $\frac{\eta}{s}\sim T^0$. On the other hand, at low
$T/k$, the hyperscaling violation emerges in the IR. The power law
of $s\sim T^\frac{d-\theta}{z}$ and $\frac{\eta}{s}\sim T^\kappa$
are observed in numerical results.

It is worthwhile to notice that $\rho/k^d$ converges to a nonzero
constant at low $T/k$. When approaching the IR, $e^2(u)$
converges to a nonzero constant for class I but goes to zero as
some power of radial coordinate $u$ for class II. Similarly,
at low $T/k$, the horizon value $e^2_h=e^2(1)$ converges to
a nonzero constant for class I but goes to zero as some
power law for class II, whose exponent is shown in Appendix
\ref{SectionIRSolution}. For the same $\{d,z,\theta\}$, it
is observed that the appearance of a nonzero $e^2$ always reduces
the exponent $\kappa$ of $\frac \eta s\sim T^\kappa$, which is
consistent with the property that $\kappa$ decreases
monotonously with $e^2$ in (\ref{bound}) when $\kappa>0$.

\begin{figure}
  \centering
  \includegraphics[height=130pt]{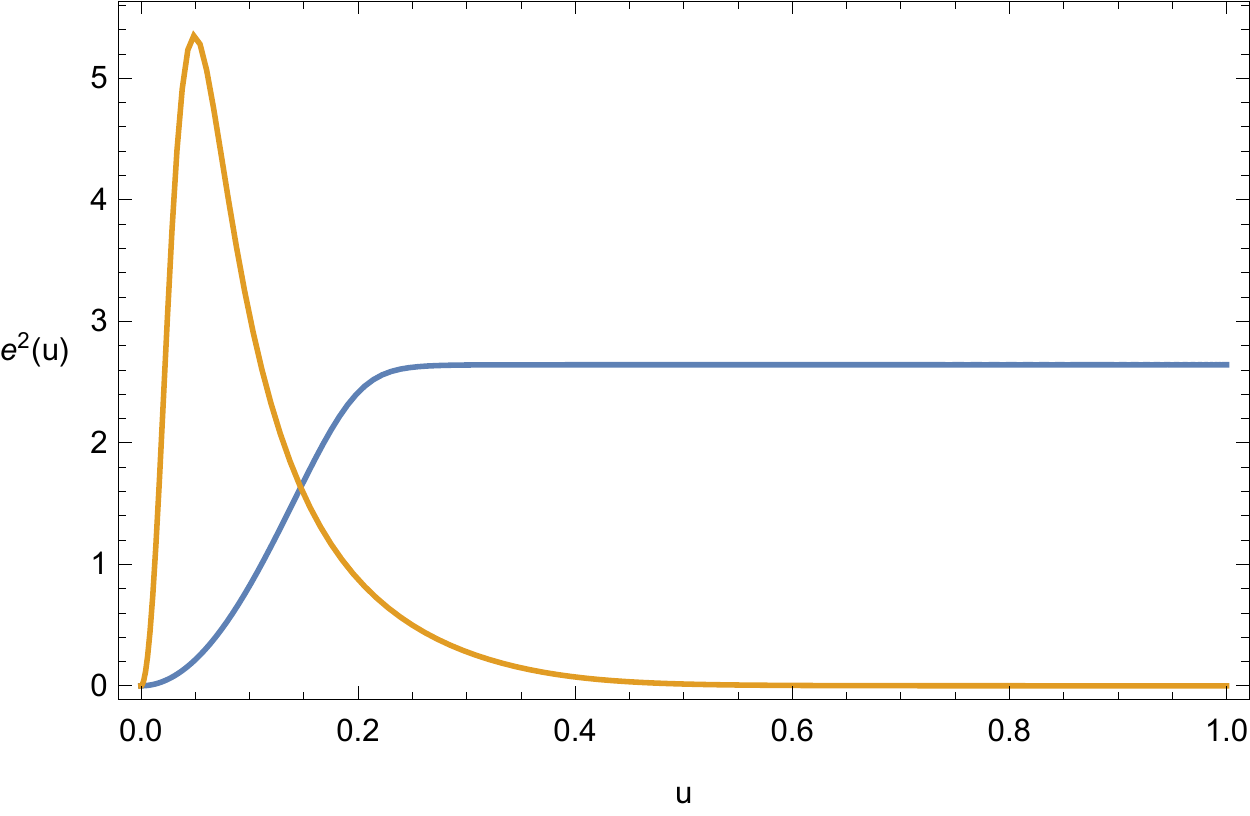}\quad
  \includegraphics[height=130pt]{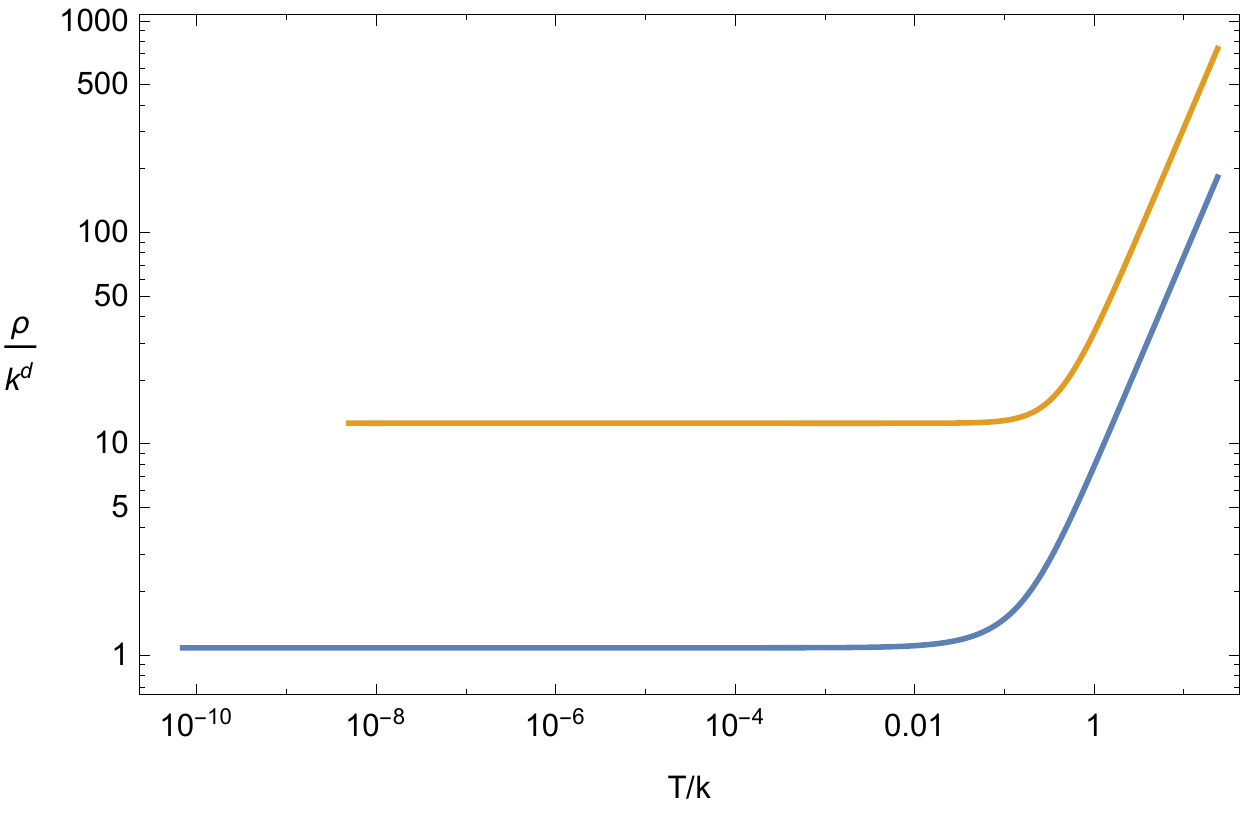}\\
  \includegraphics[height=130pt]{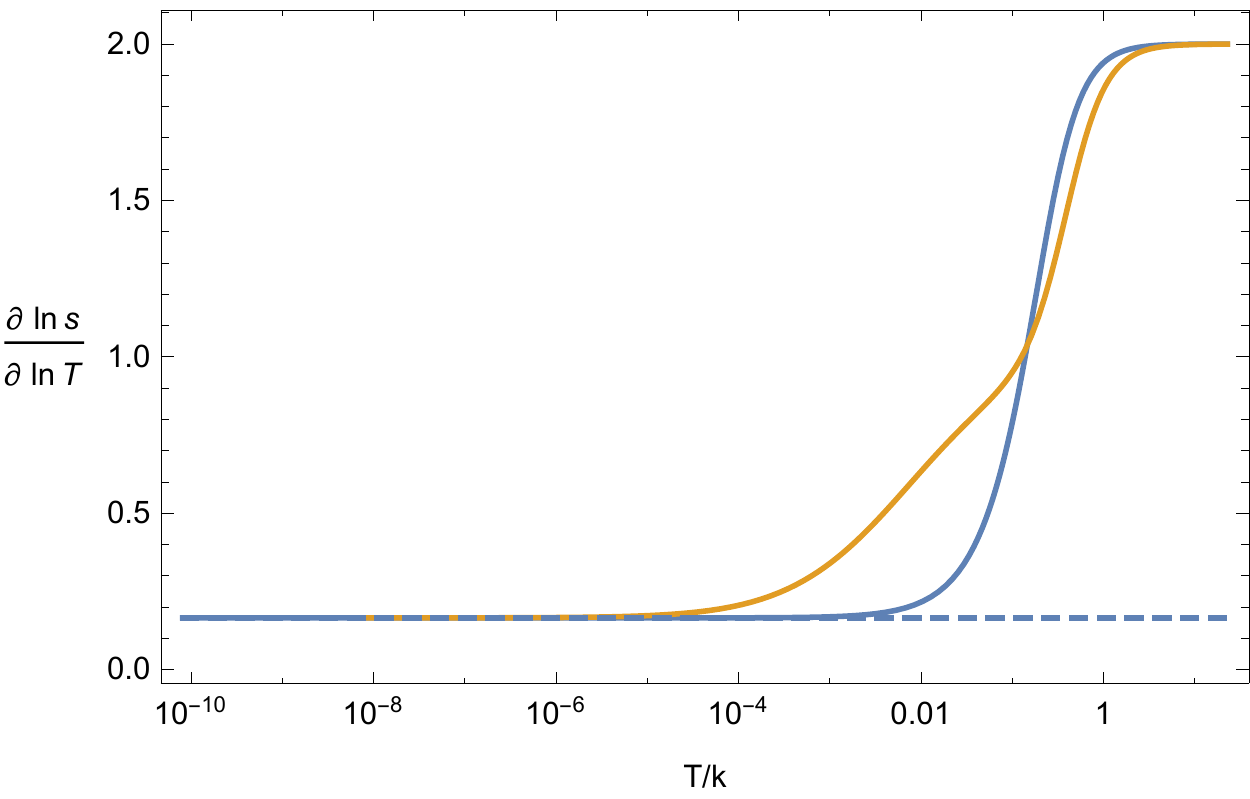}\quad
  \includegraphics[height=130pt]{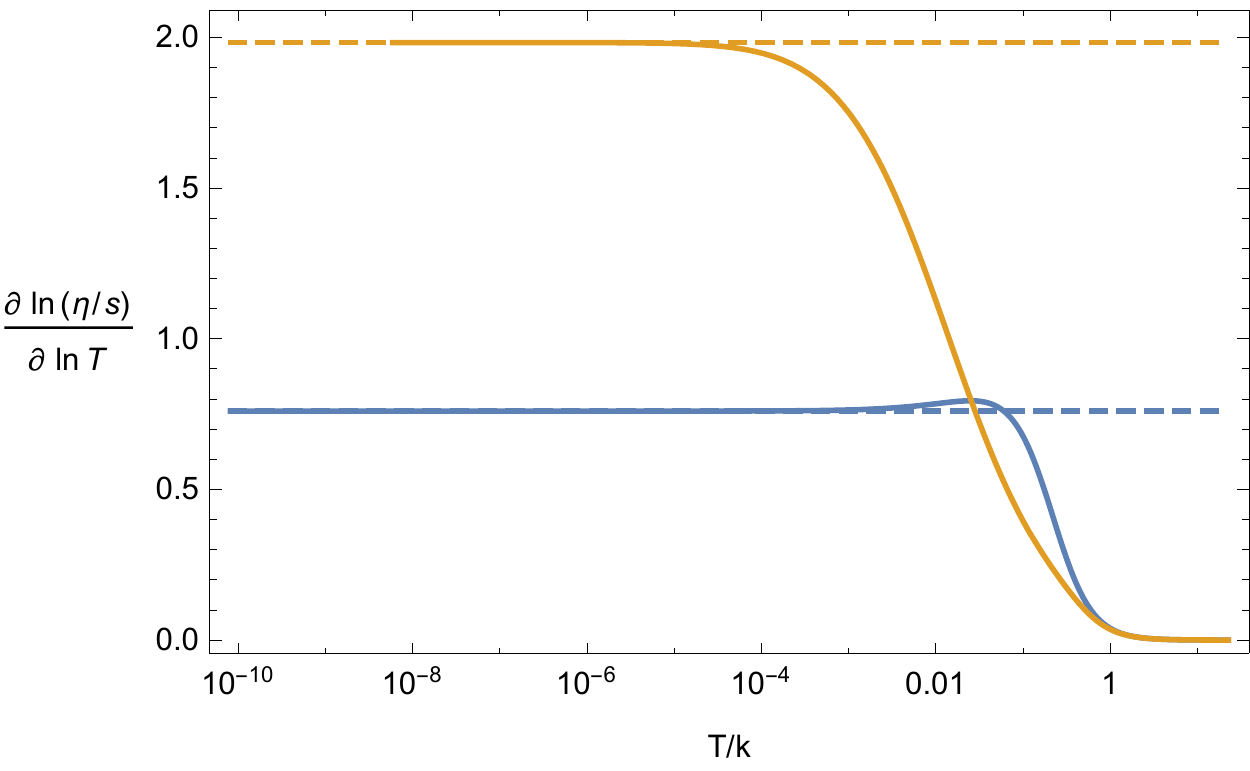}\\
  \caption{$d=2,~\theta=0$. Parameters are $\alpha=0,~\beta=-0.3$, $\gamma=0.6$ (blue) or $2$ (orange), then $z=12.1$. Among them, $\gamma=0.6$ belongs to class I; $\gamma=2$ belongs to class II. Solid lines denote numerical results and dashed lines denote analytical results from (\ref{TsHV}) and (\ref{bound}). Upper-left: Plot of $e^2(u)$ at rather low temperatures, where it converges to a nonzero constant in the IR for $\gamma=0.6$. Upper-right: Log-log plot of $\rho/k^d$ as a function of $T/k$, where $\rho/k^d$ converges to a constant at low temperature. Lower-left: Exponent $\lambda$ of $s\sim T^\lambda$ as a function of $T/k$. Lower-right: Exponent $\kappa$ of $\frac{\eta}{s}\sim T^\kappa$ as a function of $T/k$. }\label{Figd2th0}
\end{figure}

\begin{figure}
  \centering
  \includegraphics[height=130pt]{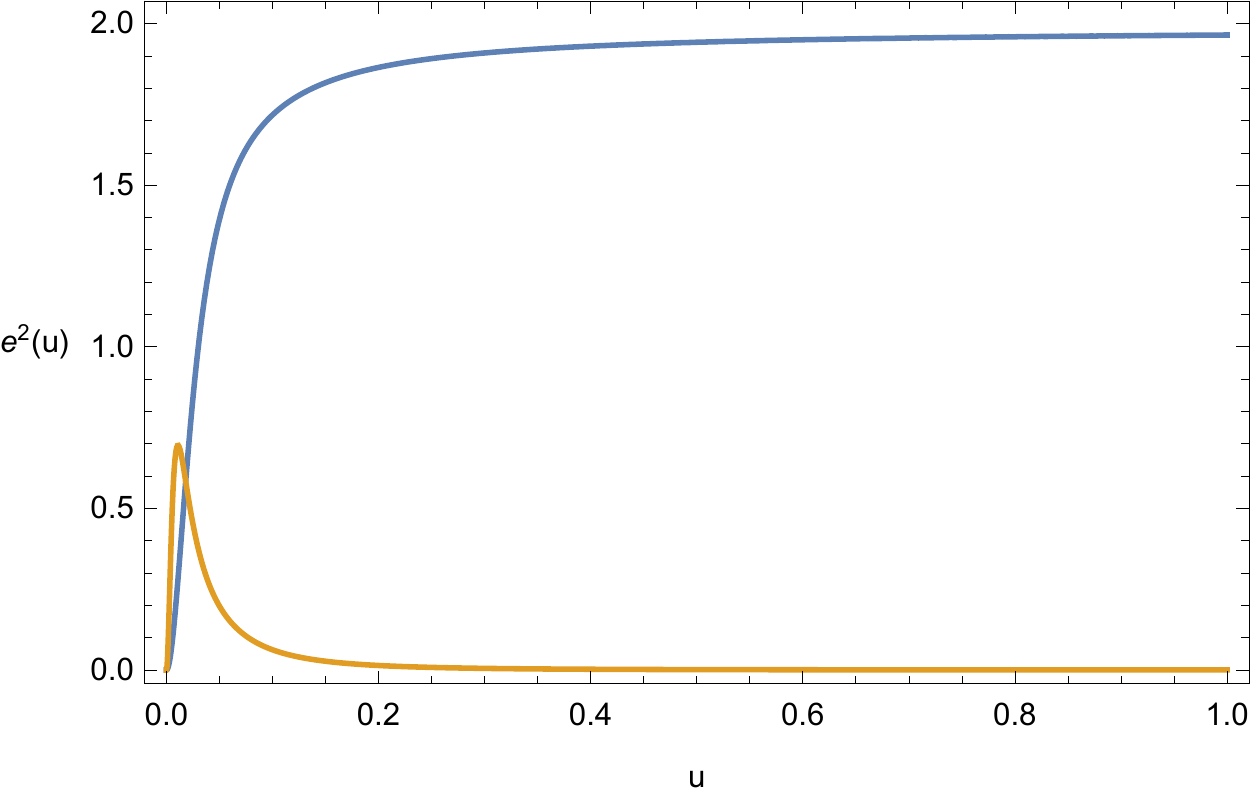}\quad
  \includegraphics[height=130pt]{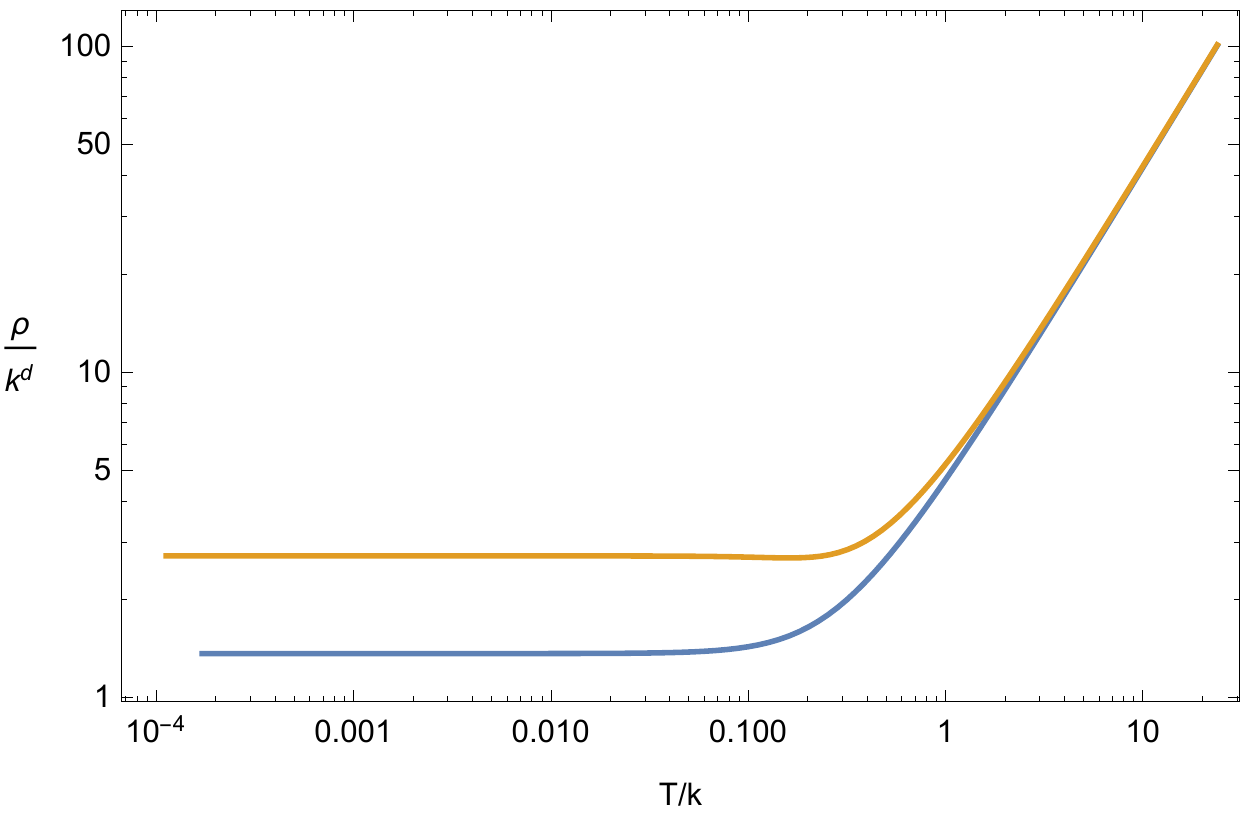}\\
  \includegraphics[height=130pt]{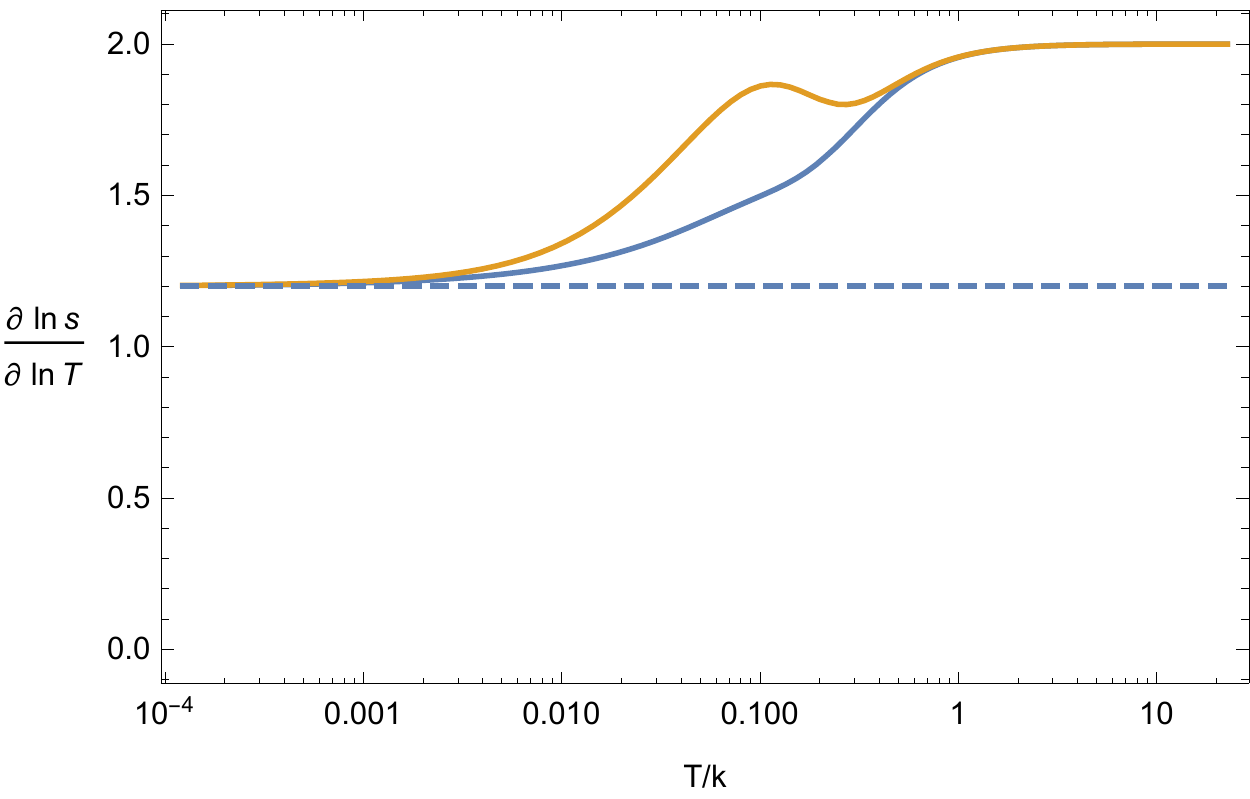}\quad
  \includegraphics[height=130pt]{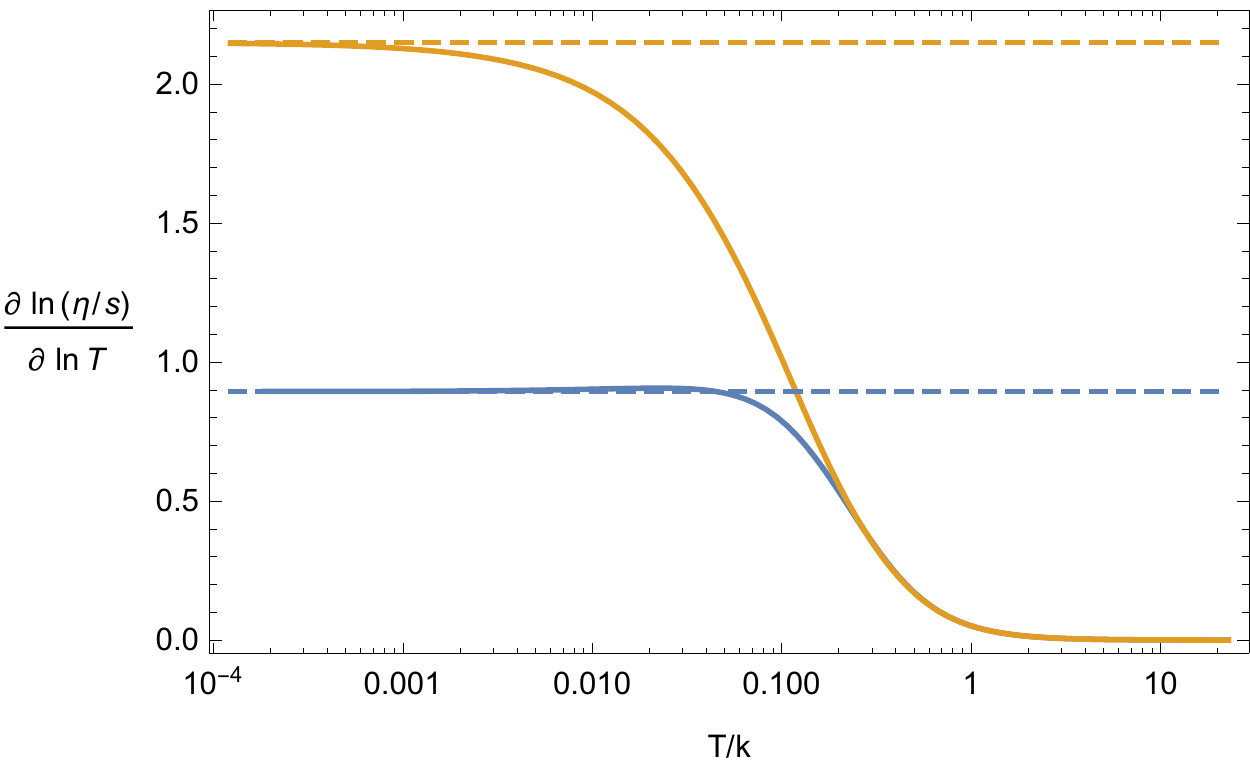}\\
  \caption{$d=2,~\theta<0$. Parameters are $\alpha=0.471,~\beta=-0.236$, $\gamma=0.943$ (blue, class I) or $2$ (orange, class II), then $z=5$ and $\theta=-4$.
  }\label{Figd2th-4}
\end{figure}

\begin{figure}
  \centering
  \includegraphics[height=130pt]{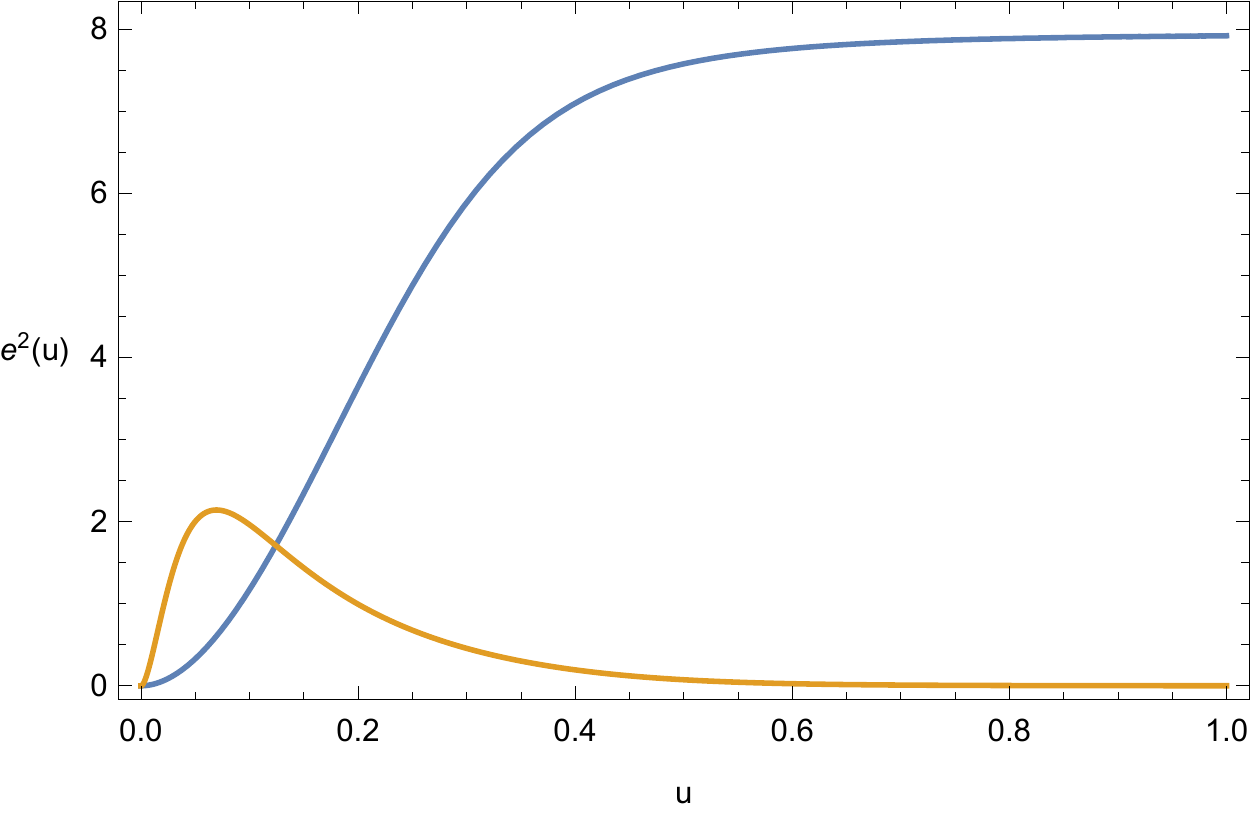}\quad
  \includegraphics[height=130pt]{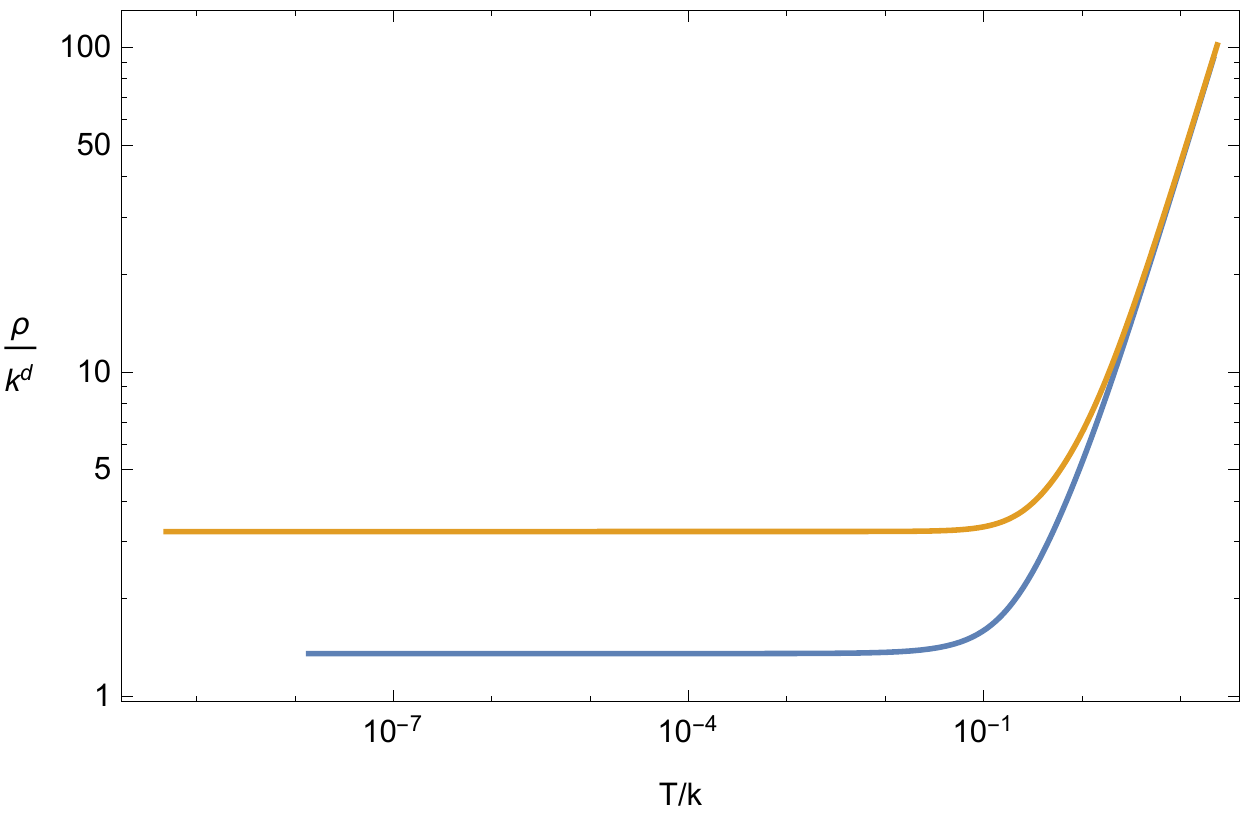}\\
  \includegraphics[height=130pt]{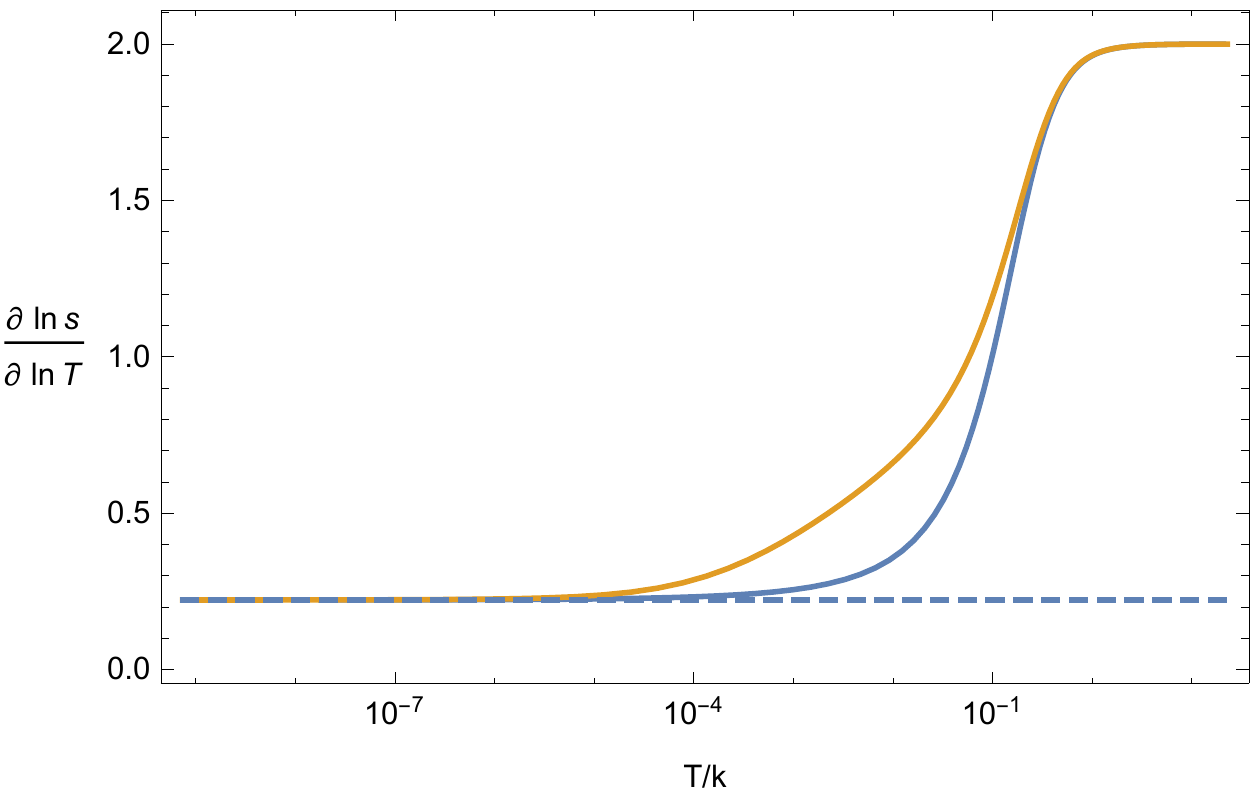}\quad
  \includegraphics[height=130pt]{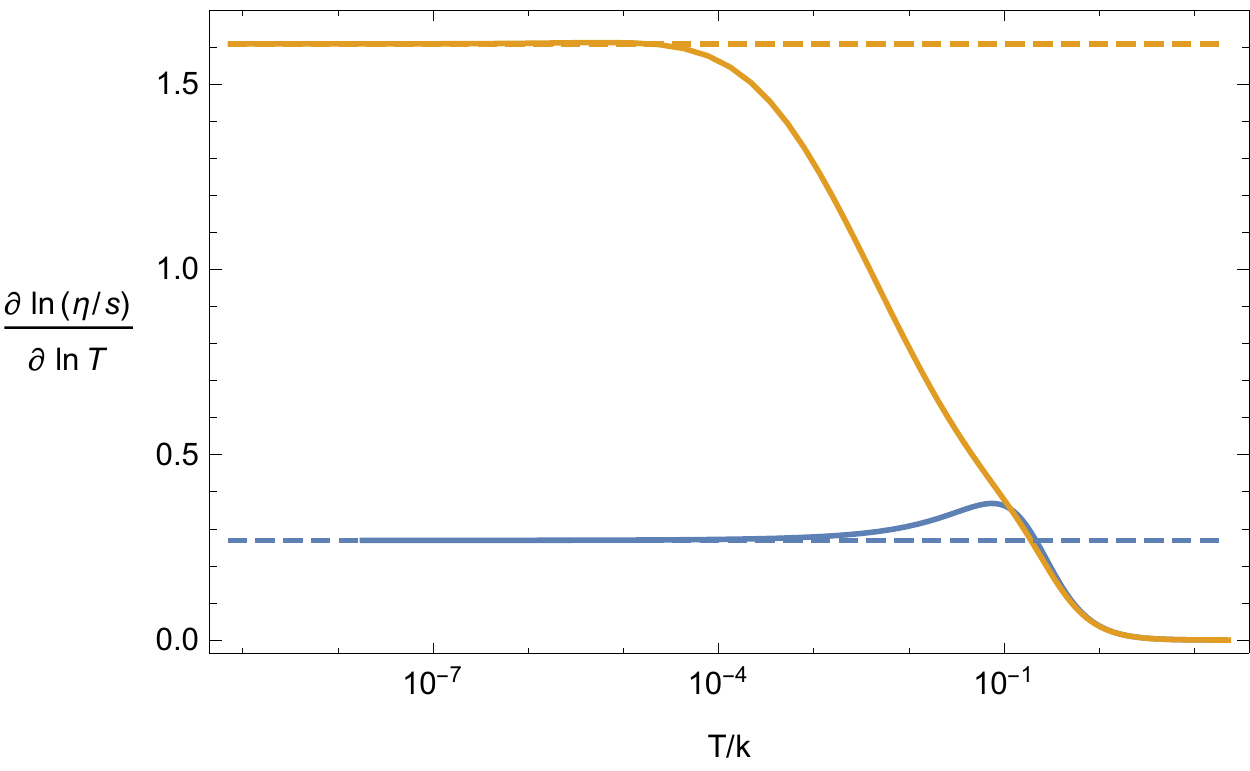}\\
  \caption{$d=2,~\theta>0$. Parameters are $\alpha=-1,~\beta=-1.5$, $\gamma=2$ (blue, class I) or $4$ (orange, class II), then $z=3$ and $\theta=1.33$.
  }\label{Figd2th1p33}
\end{figure}

We conduct the numerical calculation for $d=2$ and
$\theta=\frac{4}{3},~0,~-4$ as representives of three cases,
namely $\theta<0,~\theta=0$ and $0<\theta<d$. Different
values of $\gamma$ are chosen to represent class I or class II.
The specific results are shown in Figure \ref{Figd2th0},
\ref{Figd2th-4}, \ref{Figd2th1p33}. All the numerical results
match the analytical ones.

\subsubsection{\bf Canonical ensemble}

In canonical ensemble, we control the value of charge density
$\rho$. The $t$ component of Maxwell equations (\ref{ME}) can be
replaced by (\ref{rho}).

When $\theta=0$, the three-parameter family of solutions is
characterized by $\{\frac{T}{k},~\frac{\rho}{k^d},~\lambda\}$.
When $\theta\neq0$, the one is characterized by
$\{\frac{T}{k},~\frac{\rho}{k^d},~\frac{\lambda}{k}\}$. When
lowering down $T/k$, we fix the other two dimensionless
parameters within an appropriate region. Then $\mu/k$ converges to a nonzero constant at
low $T/k$ instead of $\rho/k^d$ in grand canonical ensemble. The
behaviors of $\eta/s$ and $e^2$ are similar to those in grand
canonical ensemble. By using the method in Appendix
\ref{SectionIRSolution}, we can foresee the value to which $e^2$
converges in the IR at low temperature for class I.

\begin{figure}
  \centering
  \includegraphics[height=130pt]{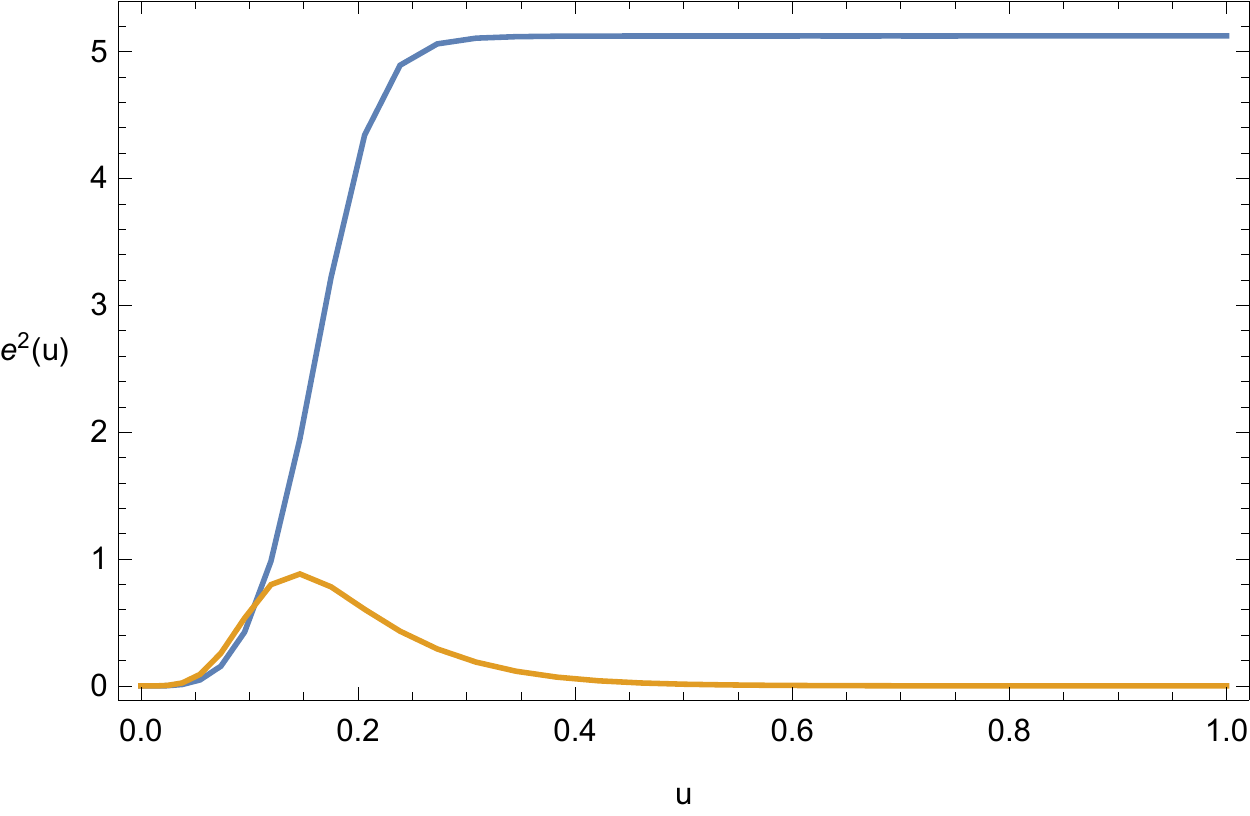}\quad
  \includegraphics[height=130pt]{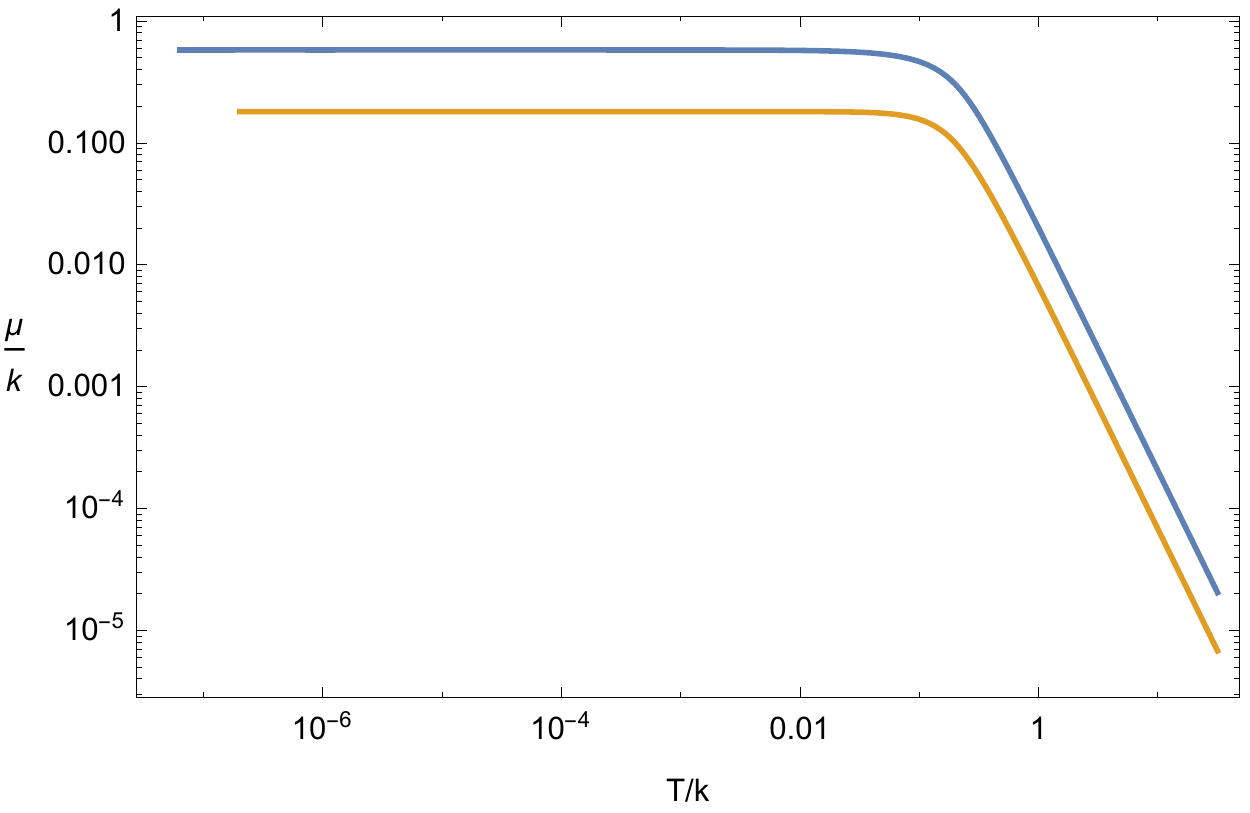}\\
  \includegraphics[height=130pt]{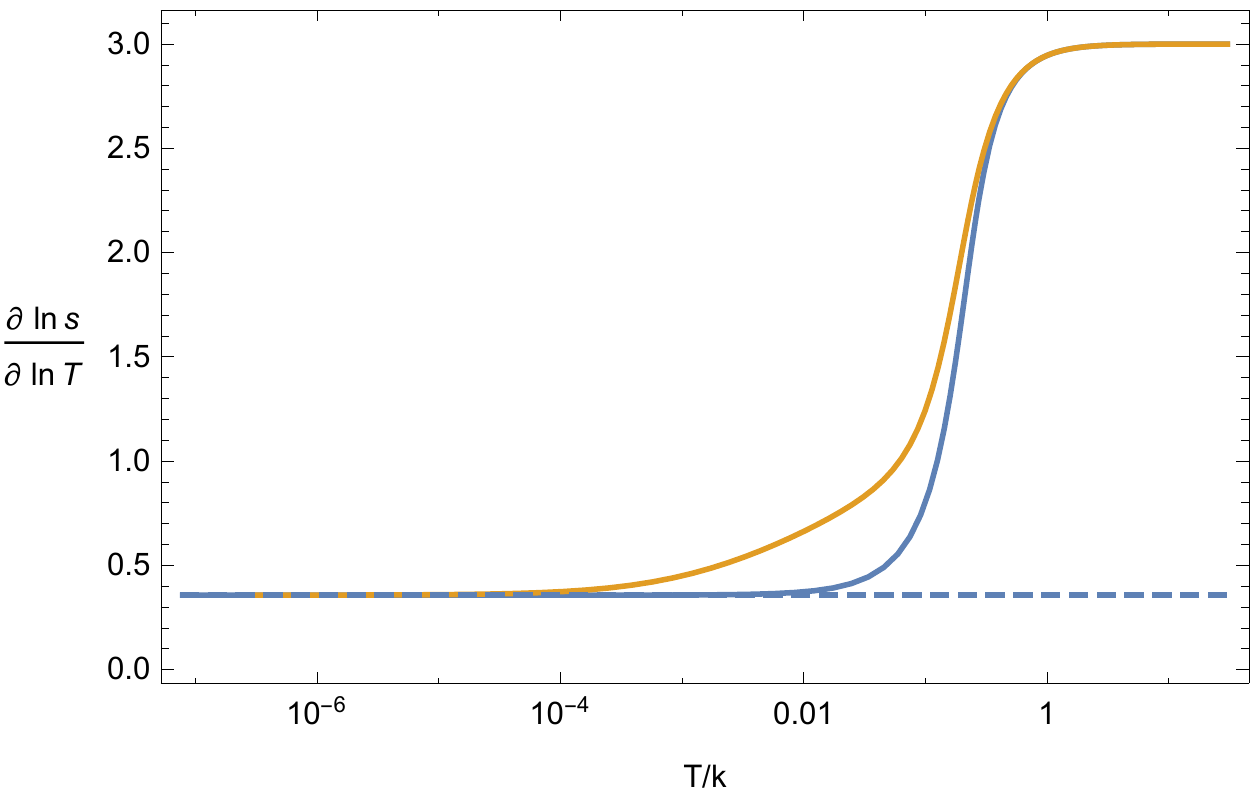}\quad
  \includegraphics[height=130pt]{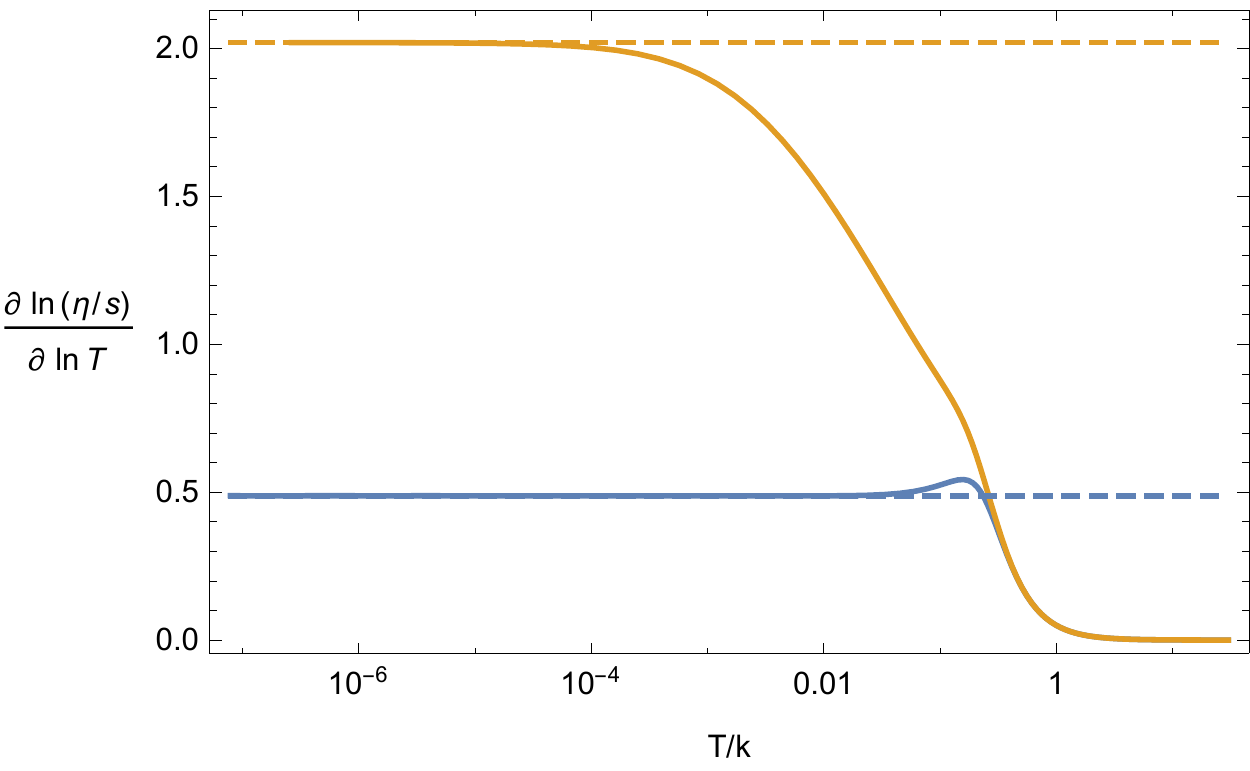}\\
  \caption{$d=3,~\theta=0$. Parameters are $\alpha=0,~\beta=-0.3$, $\gamma=0.9$ (blue, class I) or $2$ (orange, class II), then $z=8.41$. In the lower-right plot, it is worthwhile to notice that $\kappa$ converges to a quantity which is greater than $2$ at low temperature.
  }\label{Figd3th0}
\end{figure}

\begin{figure}
  \centering
  \includegraphics[height=130pt]{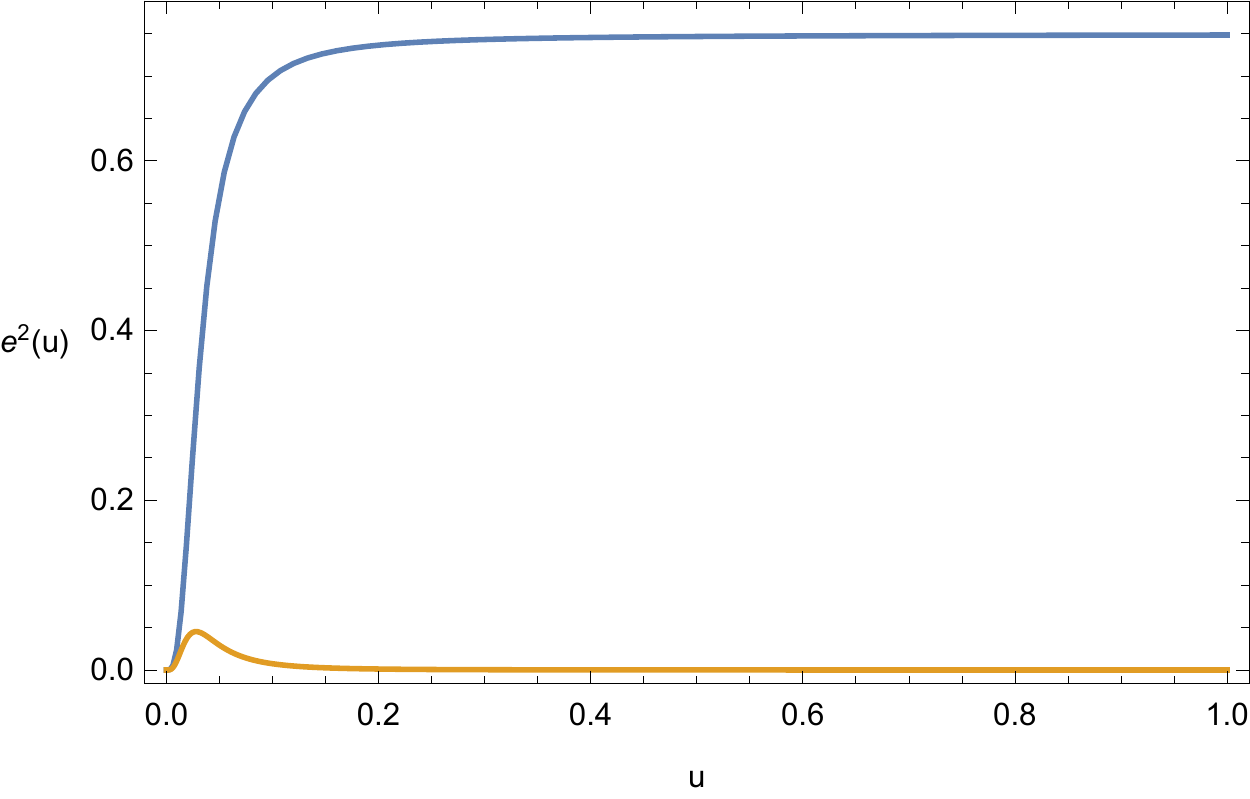}\quad
  \includegraphics[height=130pt]{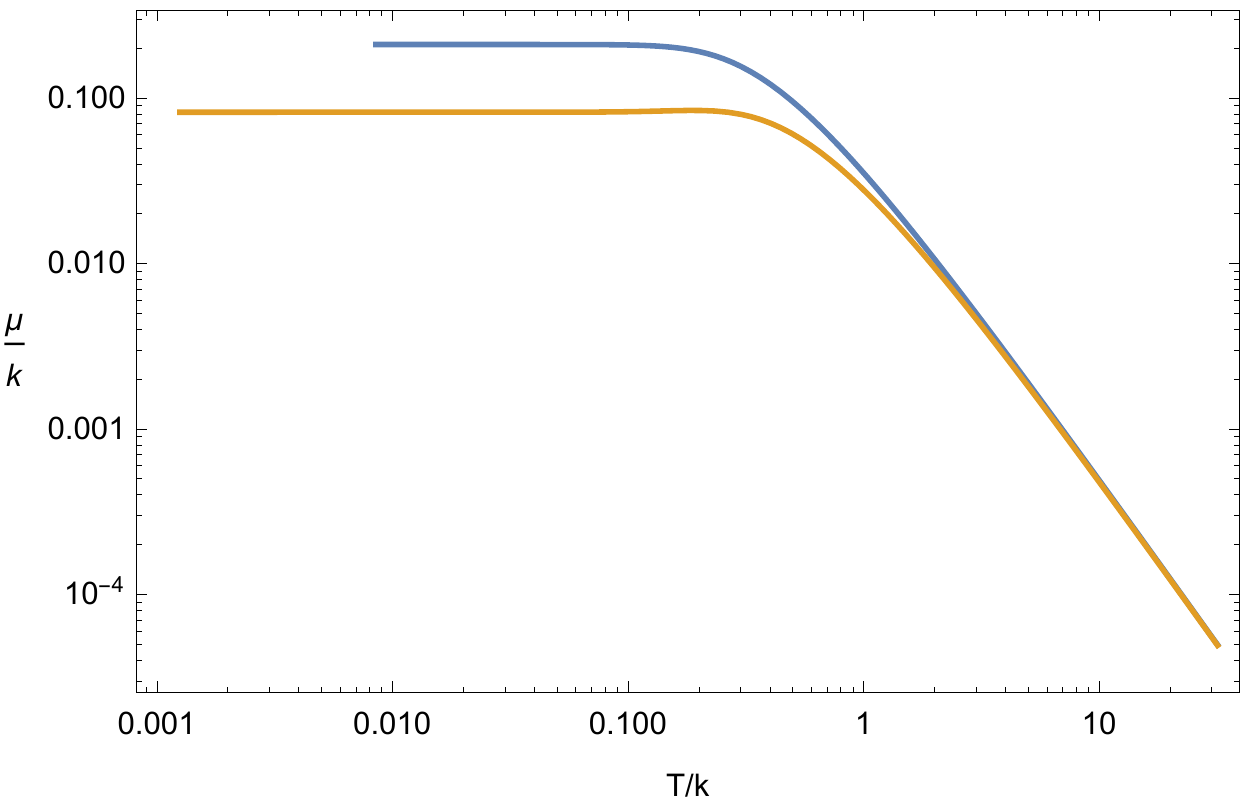}\\
  \includegraphics[height=130pt]{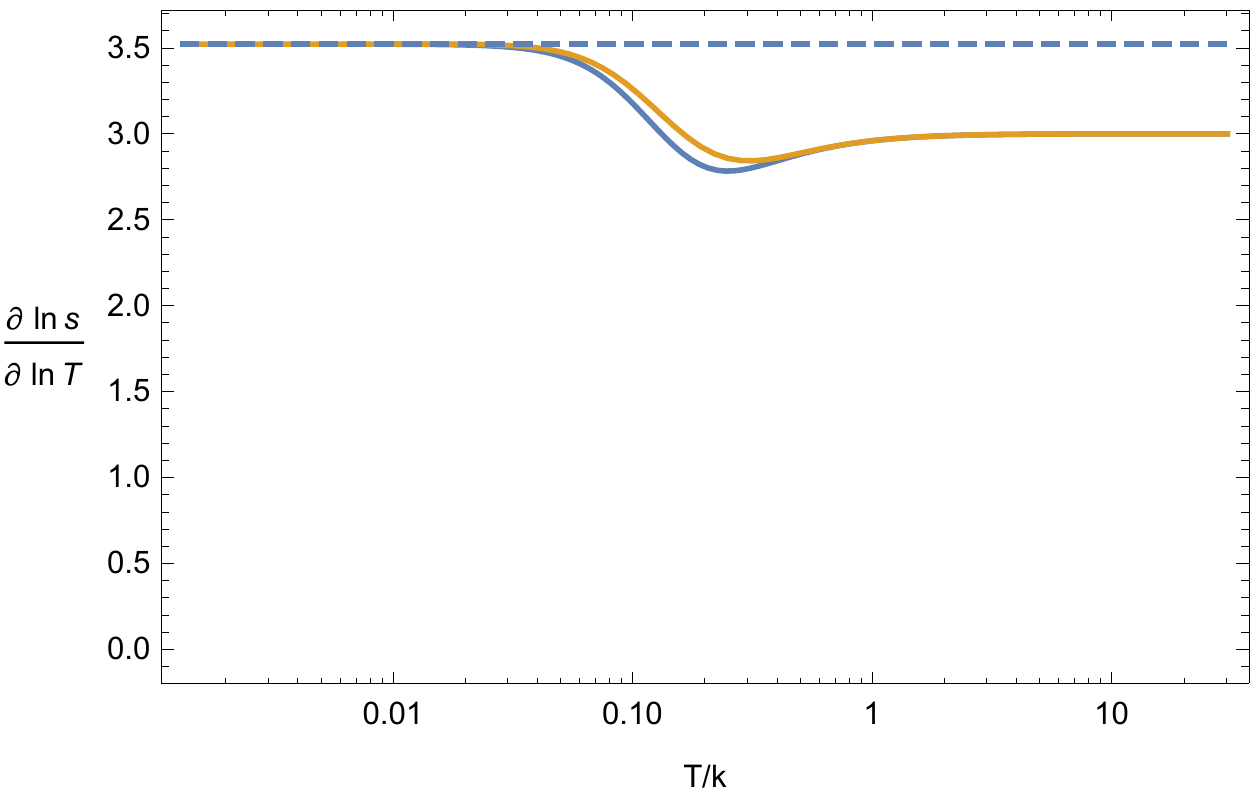}\quad
  \includegraphics[height=130pt]{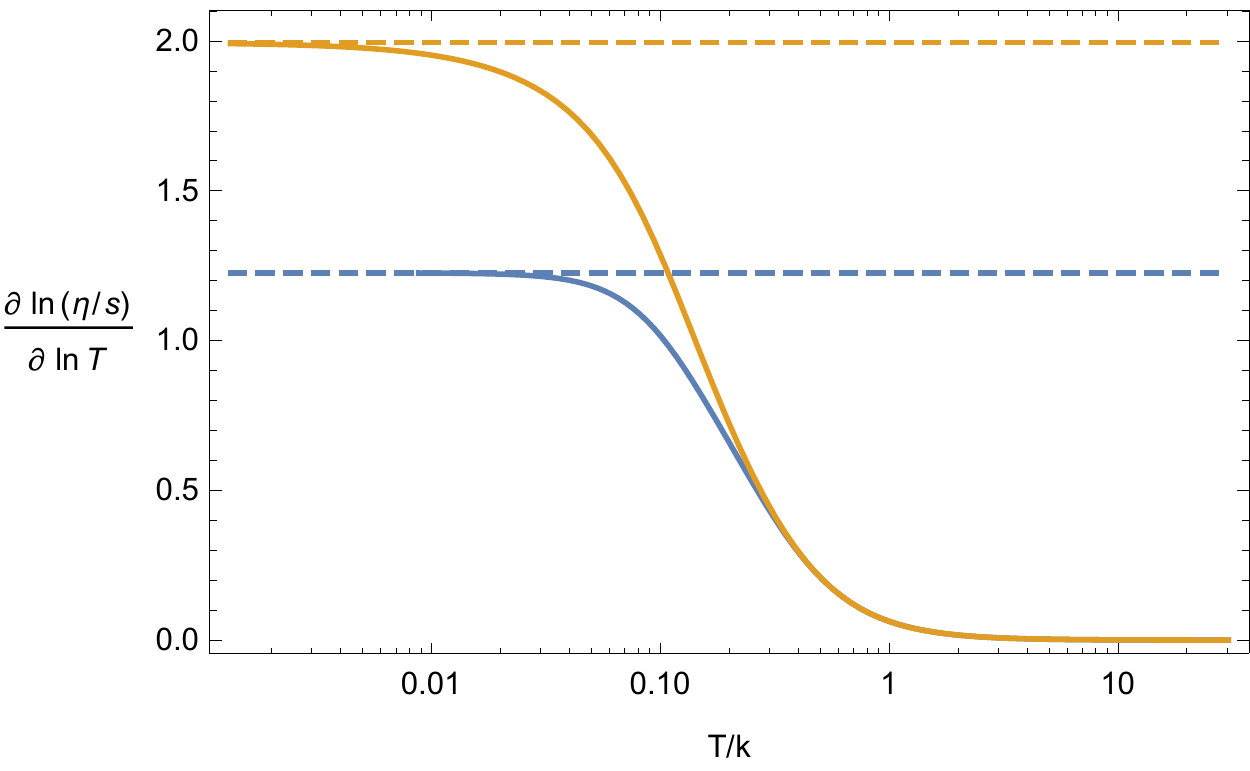}\\
  \caption{$d=3,~\theta<0$. Parameters are $\alpha=0.5,~\beta=-0.25$, $\gamma=1.75$ (blue, class I) or $3$ (orange, class II), then $z=2.56$ and $\theta=-6$.
  }\label{Figd3th-6}
\end{figure}

\begin{figure}
  \centering
  \includegraphics[height=130pt]{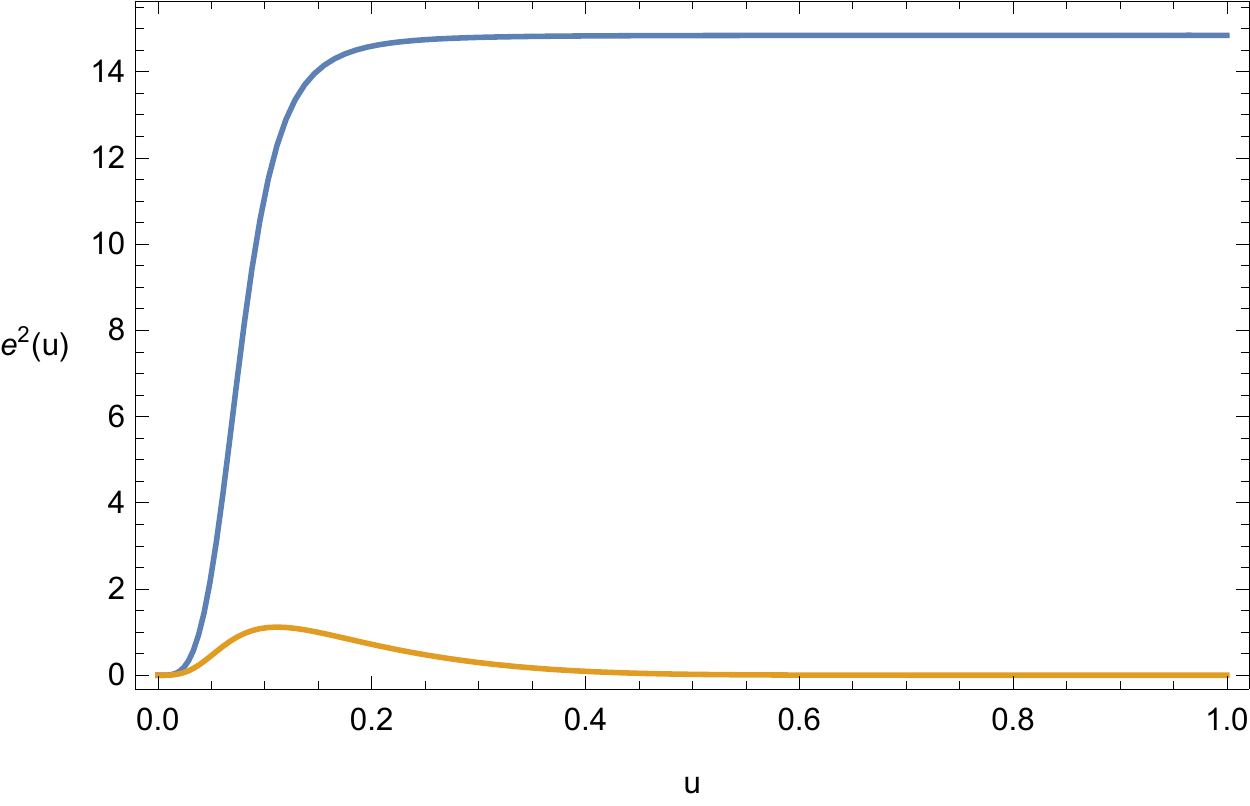}\quad
  \includegraphics[height=130pt]{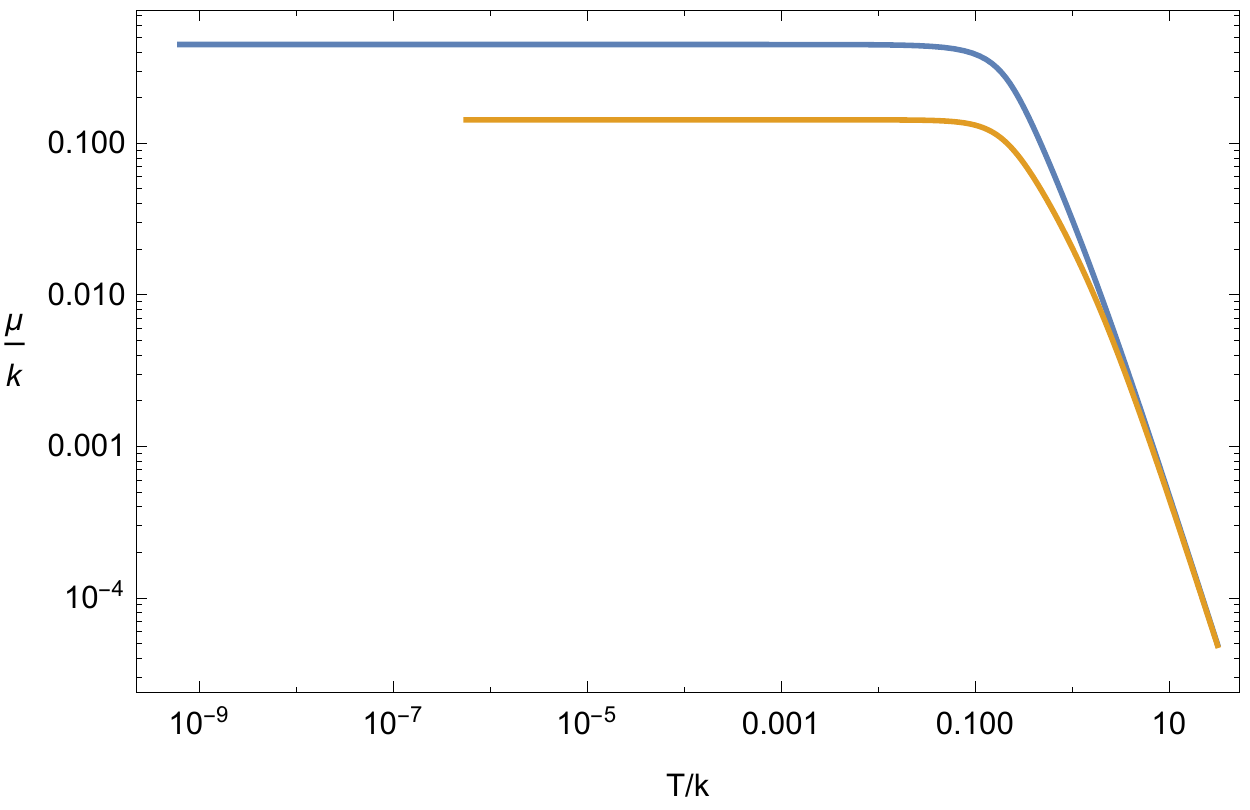}\\
  \includegraphics[height=130pt]{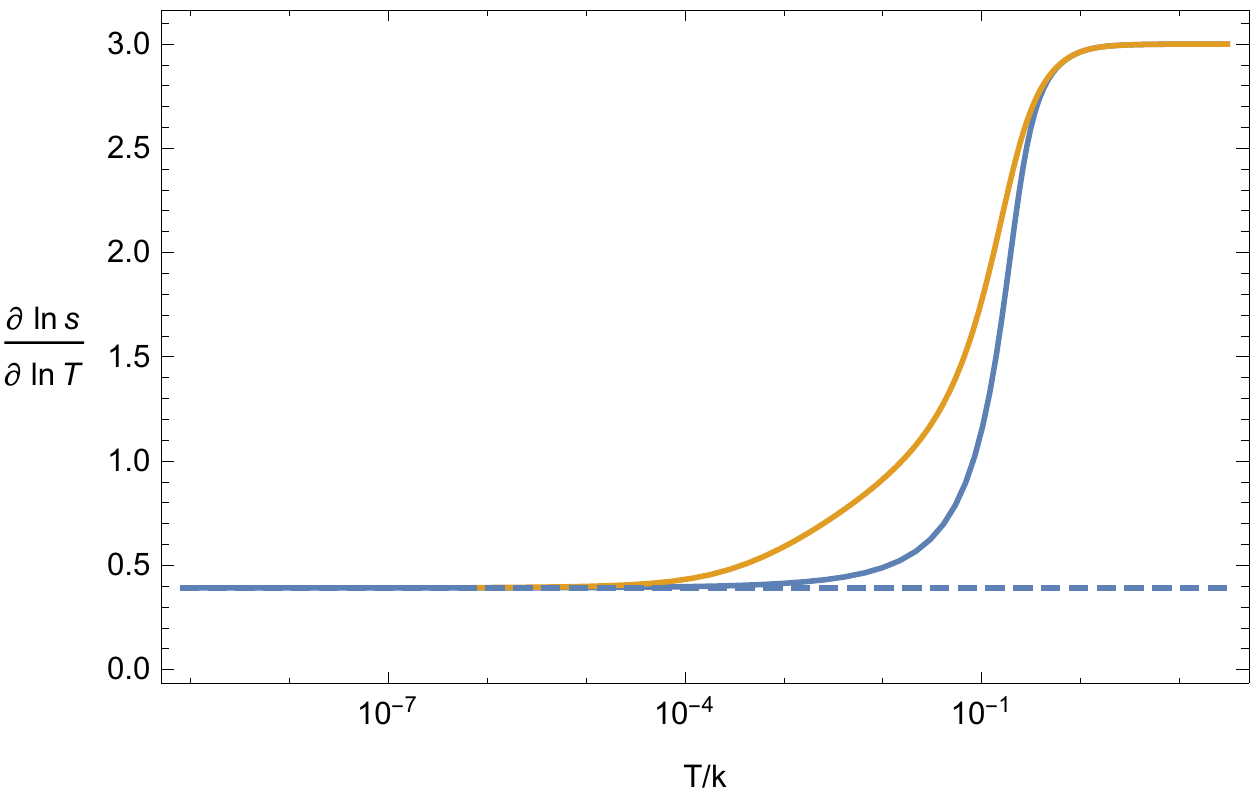}\quad
  \includegraphics[height=130pt]{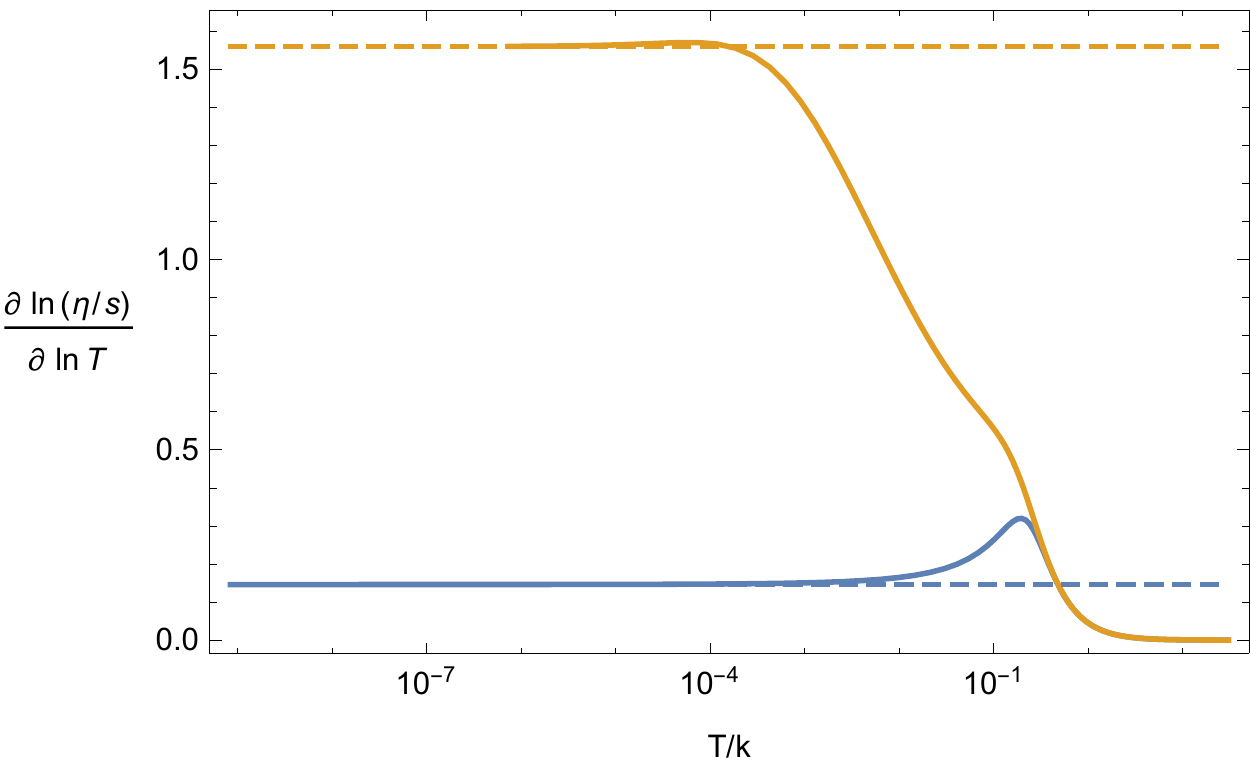}\\
  \caption{$d=3,~\theta>0$. Parameters are $\alpha=-1,~\beta=-1.5$, $\gamma=2.5$ (blue, class I) or $5$ (orange, class II), then $z=2.56$ and $\theta=2$.
  }\label{Figd3th2}
\end{figure}

In a parallel manner, we conduct the numerical calculation
for $d=3$ and $\theta=-6,~0,~2$ to represent the three regions
of $\theta<0,~\theta=0$ and $0<\theta<d$. The specific
results are shown in Figure \ref{Figd3th0}, \ref{Figd3th-6} and
\ref{Figd3th2}. All the numerical results match the analytical
ones.

\section{Conclusion and Outlooks}

In this paper we have numerically constructed charged solutions
with emerging hyperscaling violation in EMD-Axion model
and investigated the temperature behavior of the ratio of shear
viscosity to entropy density. We have found that the relevant
axion, which breaks the translational invariance, leads to the
power law of $\eta/s\sim T^\kappa$. In particular, the relevant
current reduces the exponent $\kappa$ indeed. This reduction is
characterized by the quantity $e^2$, which can be derived from
dimensionless conserved charge density $\rho/k^d$. While,
irrelevant current does not affect the exponent $\kappa$ since
$e^2\to0$ in the far IR at low temperature. Our analytical results
for the exponent $\kappa$ coincide with our numerical calculation,
indicating that our proposed formula for $\kappa$ in
(\ref{scaleformula}) is robust at least for generic backgrounds
within EMD-Axion model.

Especially, our results for Lifshitz case verify that $\kappa>2$
can happen in $d>2$, indicating that hyperscaling violation is not
the essential ingredient leading to the exponent $\kappa>2$.
Moreover, as conjectured in \cite{Ling:2016ien}, the upper bounds
for $\kappa$ coincide with the behaviors of entanglement entropy.

Analytically, it is possible that irrelevant current or axion can
affect the temperature behavior of $\eta/s$ at subleading
orders. One should consider their backreaction to the
background, then solve the shear perturbation equation
(\ref{perturh}). It is related to the issue whether
temperature $T$ is still the unique scale in entropy
production, as it is when axion is relevant. It is worthy of
investigation in the future.

\acknowledgments{
We are very grateful to Blaise Gout\'eraux, Peng Liu and Xiangrong
Zheng for helpful discussions and correspondence.}

\begin{appendix}

\section{The classification of IR solutions}\label{SectionIRSolution}
We focus on relevant axion solutions, otherwise $\eta/s$ just
converges to a non-zero constant at low temperature at
leading order. We require the solutions should have positive
specific heat, and the temperature deformation is the only
allowed relevant deformation. In addition, we give an extra
requirement of $\theta<d$. The scaling solutions have been
obtained and classified in \cite{Gouteraux:2014hca,Donos:2014uba}.

\subsection{Class I: marginally relevant current, marginally relevant axion}
If both the Maxwell and the axion terms are the same order in the
power of the radial coordinate as the curvature term and the dilaton
potential $V(\phi)$ in Lagrange (\ref{action}), scaling solutions
obtained form a one-parameter family
\begin{equation}\begin{split}
&
\beta  \epsilon =-2,\quad \alpha  \epsilon =-\frac{2 \theta }{d},\quad\gamma =\alpha  (d-1)-\beta  d,\quad \epsilon ^2=\frac{2 (d-\theta ) (d (z-1)-\theta )}{d},
\\&
\zeta=\theta-d,\quad L^2=\frac{2 \left(\delta _0-1\right) \delta _0}{2 V_0-(d-1) k^2},\quad Q^2=\frac{2 \left(k^2 (d z-\theta )+2 V_0 (1-z)\right)}{\delta _0 \left((d-1) k^2-2 V_0\right)},
\end{split}\end{equation}
which can be parameterized by $k$ in coordinate of (\ref{HV}). The charge related quantities are
\begin{eqnarray}
\rho^2&=&\frac{\delta_0^2Q^2}{L^2}=\frac{k^2 (\theta -d z)+2 V_0 (z-1)}{\delta _0-1}, \label{rhoclassI} \\
e^2&=&\frac{\delta_0^2Q^2}{L^2 k^2}=\frac{k^2 (\theta -d z)+2 V_0 (z-1)}{ k^2 (\delta_0-1)}. \label{e2classI}
\end{eqnarray}

The mode analysis in \cite{Gouteraux:2014hca} indicates that there
are three pairs of conjugate modes summing to $\delta_0$. Two
pairs are degenerate with $\beta_{1,-}=\beta_{2,-}=0$ and
$\beta_{1,+}=\beta_{2,+}=\delta_0$. $\beta_{1,-}$ rescales the
time. $\beta_{1,+}$ is temperature deformation and responsible for
creating a small black hole (\ref{HVBH}). $\beta_{2,-}$ changes
$k$ and shifts the solution along the one-parameter family.
$\beta_{2,+}$ changes the chemical potential and belongs to the
transformation of gauge symmetry. The expression for the
last pair of $\beta_{3,\pm}$ is too tedious to show here. We
require that $\beta_{1,+}$ is relevant and $\beta_{3,-}$ is
irrelevant, with IR located at $r\to\infty$. Then the final
allowed parameter space here is found to be $\rho^2>0$ and
(\ref{constraint}).

Since the quantity
$\frac{\rho}{k^d}=\frac{\sqrt{-g_{tt}g_{rr}g_{xx}^d}Z(\phi)F^{rt}}{(\partial_x\chi)^d}$
is conserved and invariant under coordinate transformation within (\ref{metricandemtensor}), we
can use it to connect UV with IR and determine the solution in the
one-parameter family by using (\ref{rhoclassI}). Finally, $e^2$
can be obtained from $\frac{\rho}{k^d}$ by using (\ref{e2classI}),
which is rather convenient in canonical ensemble.

At zero temperature, one can integrate the three modes of
$\beta_{2,-}$, $\beta_{2,+}$ and $\beta_{3,-}$ to the UV and
adjust them to satisfy the boundary conditions specified by
$\{\lambda,\rho,\mu\}$ at the conformal boundary. A finite
temperature solution can be driven by $\beta_{1,+}$.

\subsection{Class II: irrelevant current, marginally relevant axion}
If only the Maxwell term turns to be subleading, a single scaling
solution at leading order is obtained as
\begin{equation}\begin{split}
&
\beta \epsilon =-2,\quad \alpha  \epsilon =-\frac{2 \theta }{d}, \quad \epsilon ^2=\frac{2 (d-\theta ) (d (z-1)-\theta )}{d},
\\&
L^2=\frac{\delta _0 (d z-\theta )}{V_0}, \quad k^2=\frac{2 V_0 (z-1)}{d z-\theta }.
\end{split}\end{equation}
There are three pairs of modes, in which two pairs sum to
$\delta_0$. The first pair is $\beta_{1,-}=0$ and
$\beta_{1,+}=\delta_0$ which are rescaling of time and
temperature deformation. The second pair is relevant $\beta_{2,+}$
and irrelevant $\beta_{2,-}$. The third pair is gauge field modes
with
\begin{equation}
A(r)= A_1 + A_2 r^{\zeta-z}, \zeta = d-\theta+\frac{2\theta}{d}-\epsilon \gamma.
\end{equation}
Mode of $A_2$ is irrelevant when $\delta_0 (\zeta+d-\theta)<0$. We
find $e^2(r)\sim r^{\zeta-d+\theta}$ and $\rho \sim T^0, e^2_h
\sim T^\frac{\zeta-d+\theta}{-z}$, when $T\to 0$. Similar to class I, we can integrate
$\beta_{2,-}$ and gauge field modes to the UV.

\end{appendix}

\centerline{\rule{80mm}{0.1pt}}

\end{document}